\documentclass[aps,prl,twocolumn,amsfonts]{revtex4-1}

\usepackage{graphicx}
\usepackage{amsmath}
\usepackage{relsize}

\usepackage[colorlinks=true,citecolor=blue,linkcolor=magenta]{hyperref}
\hypersetup{
pdfauthor = {A. Barfuss},
colorlinks = true, linkcolor = blue, urlcolor=blue, bookmarksnumbered =  true}

\newcommand{\ket}[1]{\left|#1\right>}

\newcommand*\myPMbP{\ensuremath{\substack{\mathsmaller(+\mathsmaller)\\[-0.25em]-}\,}}

\begin{document}

\title{Non-reciprocal coherent dynamics of a single spin under closed-contour interaction}

\author{Arne Barfuss}
\affiliation{Department of Physics, University of Basel, Klingelbergstrasse 82, CH-4056 Basel, Switzerland}
\author{Johannes K\"olbl}
\affiliation{Department of Physics, University of Basel, Klingelbergstrasse 82, CH-4056 Basel, Switzerland}
\author{Lucas Thiel}
\affiliation{Department of Physics, University of Basel, Klingelbergstrasse 82, CH-4056 Basel, Switzerland}
\author{Jean Teissier}
\affiliation{Department of Physics, University of Basel, Klingelbergstrasse 82, CH-4056 Basel, Switzerland}
\author{Mark Kasperczyk}
\affiliation{Department of Physics, University of Basel, Klingelbergstrasse 82, CH-4056 Basel, Switzerland}
\author{Patrick Maletinsky}
\affiliation{Department of Physics, University of Basel, Klingelbergstrasse 82, CH-4056 Basel, Switzerland}
\email{patrick.maletinsky@unibas.ch}

\date{\today}

\begin{abstract}
Three-level quantum systems have formed a cornerstone of quantum optics since the discovery of coherent population trapping (CPT)\,\cite{Gray1978, Arimondo1996, Agapev1993} and electromagnetically induced transparency\,\cite{Fleischhauer2005}.
Key to these phenomena is quantum interference, which arises if two of the three available transitions are coherently driven at well-controlled amplitudes and phases. 
The additional coherent driving of the third available transition would form a closed-contour interaction (CCI) from which fundamentally new phenomena would emerge, including phase-controlled CPT\,\cite{Kosachiov1992, Kosachiov1991, Windholz2001} and one atom interferometry\,\cite{Buckle1986}. 
However, due to the difficulty in experimentally realising a fully coherent CCI, such aspects of three-level systems remain unexplored as of now. 
Here, we exploit recently developed methods for coherent driving of single Nitrogen-Vacancy (NV) electronic spins to implement highly coherent CCI driving.  
Our experiments reveal phase-controlled, single spin quantum interference fringes, reminiscent of electron dynamics on a triangular lattice, with the driving field phases playing the role of a synthetic magnetic flux\,\cite{Roushan2017}. 
We find that for suitable values of this phase, CCI driving leads to efficient coherence protection of the NV spin, yielding a nearly two orders of magnitude improvement of the coherence time, even for moderate drive strengths $\lesssim1~$MHz.
Our results establish CCI driving as a novel paradigm in coherent control of few-level systems that offers attractive perspectives for applications in quantum sensing or quantum information processing.
\end{abstract}

\maketitle

\begin{figure}
	\centering
		\includegraphics{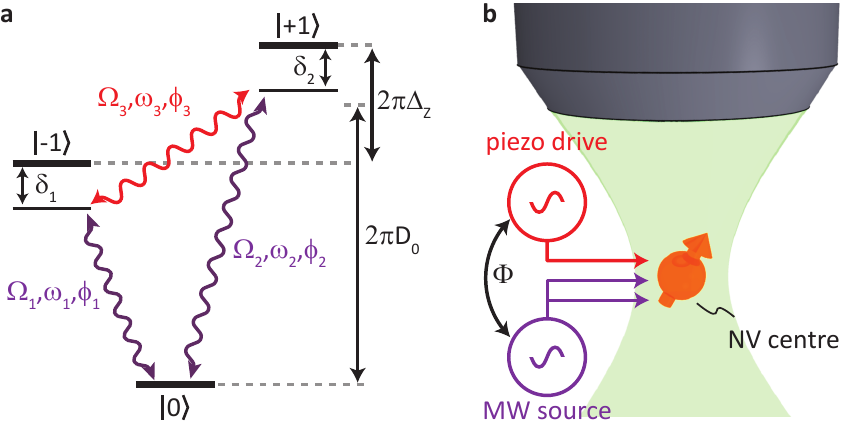}
	\caption{{\bf Closed contour interaction (CCI) scheme and experimental setup}. a) Schematic representation of the three-level CCI system studied here. Levels $\ket{-1}$, $\ket{0}$ and $\ket{+1}$ are the $S = 1$ ground state spin levels of a negatively charged diamond Nitrogen-Vacancy (NV) centre.
All three possible spin-transitions are coherently driven, either by microwave magnetic fields (for $|\Delta m_s|=1$; purple arrows) or by strain  ($|\Delta m_s|=2$; red arrow). The fields have frequency $\omega_i$, amplitude (Rabi frequency) $\Omega_i$, and phase $\phi_i$ ($i\in\{1,2,3\}$). 
b) Spin dynamics under CCI are investigated using a confocal microscope for optical initialisation and readout of the NV spin. 
Driving fields for microwave- and strain-driving are applied through a microwave antenna and piezo-excitation using appropriate generators (see SOM), which are mutually phase-locked to control the global interaction phase $\Phi=\phi_1 + \phi_3 - \phi_2$.
}
	\label{fig:fig1}
\end{figure}

The well-established approaches for addressing three-level systems rely on simultaneous, coherent driving of two dipole-allowed transitions in the system and lead to applications ranging from light-storage\,\cite{Phillips2001} and atomic clock frequency standards\,\cite{Kitching2000, Vanier2005}, to coherent quantum control\,\cite{Bergmann1998, Cirac1997, Northup2014}.
A closing of the interaction contour in the three-level system by additional, simultaneous driving of the third available transition (Fig.\,\ref{fig:fig1}a) would be of fundamental and practical interest\,\cite{Kosachiov1992, Kosachiov1991, Windholz2001,Shore2011,Buckle1986}, but is severely complicated by selection rules that prevent this closing in most experimental systems. 
Indeed, for symmetry reasons, only two of the three available transitions can be dipole-allowed for the same type of driving field.
The use of a combination of electric and magnetic dipole transitions has been proposed as a remedy for this fundamental limitation\,\cite{Shahriar1990,Maichen1995,Yamamoto1998,Korsunsky1999}. 
However, all reported systems allowing for such driving suffer from fast dephasing compared to the rate of coherent manipulation, which prevented the experimental observation of quantum coherent CCI dynamics thus far.

In this work, we exploit the unique properties of the electronic spin of the Nitrogen-Vacancy (NV) centre in diamond to overcome these limitations and observe coherent CCI and the associated, non-reciprocal dynamics in a single three-level quantum system. 
Exquisite control of the NV centre spin degrees of freedom and innovative ways to coherently drive the NV spin allow us to implement CCI and establish the strong influence of the driving field phase on coherent spin dynamics. 
A surprising outcome is the realisation that for appropriate phases of CCI driving, the scheme allows for a significant enhancement of the NV's inhomogenous dephasing time, which we extend by nearly two orders of magnitude compared to the undriven case.

The negatively charged NV centre, a substitutional nitrogen atom next to a vacancy in the diamond lattice, forms an $S=1$ spin system (Fig.\,\ref{fig:fig1}a) in its orbital ground state.  
Conveniently, the NV spin can be initialised using optical spin pumping under green laser excitation and optically read out by virtue of its spin-dependent fluorescence\,\cite{Gruber1997}. 
The NV's spin sub-levels are $\ket{0}$ and $\ket{\pm1}$, where $\ket{m_s}$ are the eigenstates of the spin operator $\hat{S}_z$ along the NV's symmetry axis $z$ (i.e. $\hat{S}_z\ket{m_s}=m_s\ket{m_s}$).
In the absence of symmetry breaking fields, the electronic spin states $\ket{\pm1}$ are degenerate and shifted from $\ket{0}$ by a zero-field splitting $D_0=2.87~$GHz.
Applying a static magnetic field $B_\mathrm{NV}$ along $z$ splits $\ket{\pm1}$ by $\Delta_Z = 2\gamma_\mathrm{NV}B_\mathrm{NV}$, with $\gamma_\mathrm{NV} = 2.8~$MHz/G, and leads to the formation of the three-level ``$\nabla$-system" we study in this work (Fig.\,\ref{fig:fig1}a)\,\cite{Doherty2013}. 
Although each of the states $\ket{m_s}$ has additional nuclear degrees of freedom due to hyperfine coupling to the NV's $^{14}$N nuclear spin, we  restrict ourselves to the hyperfine sub-space with nuclear spin quantum number $m_I= +1$\,\cite{Barfuss2015}, while other states remain out of resonance with our driving fields and do not contribute to CCI dynamics. 

To implement and study CCI dynamics, we employ coherent driving of the NV spin using a combination of time-varying magnetic and strain fields. Specifically, we use the well established method of coherent driving of the $\ket{0}\leftrightarrow\ket{\pm1}$ transitions with microwave magnetic fields\,\cite{Dobrovitski2013}. In addition, we utilise a time-varying strain field to drive the $\ket{-1}\leftrightarrow\ket{+1}$ transition -- a recently developed method for efficient, coherent driving of this magnetic dipole-forbidden transition, which is difficult to address otherwise\,\cite{Barfuss2015,MacQuarrie2015}. Considering the combined action of these three driving fields of amplitudes (Rabi frequencies) $\Omega_i$ and frequencies $\omega_i$ (Fig.\,\ref{fig:fig1}a), the dynamics of the NV spin in an appropriate rotating frame\,\cite{SOM} are described by the Hamiltonian
\begin{equation}
	\hat{\boldsymbol{H}}_0 = \frac{\hbar}{2}
	\begin{pmatrix} 
	2 \delta_1 & \Omega_1 & \Omega_3 e^{i\Phi} \\
	\Omega_1 &  0 & \Omega_2 \\ 
	\Omega_3 e^{-i\Phi} & \Omega_2 & 2 \delta_2
	\end{pmatrix},
	\label{equ:HamiltonianRWA}
\end{equation}
if the three-photon resonance $\omega_1+\omega_3=\omega_2$ is fulfilled ($\hbar$ is the reduced Planck constant). Hamiltonian $\hat{\boldsymbol{H}}_0$ is expressed in the basis $\{\ket{-1}, \ket{0}, \ket{+1}\}$ and $\delta_{1(2)}$ represent the detunings of the microwave driving fields from the $\ket{0}\leftrightarrow\ket{ \myPMbP 1}$ spin transition.
Importantly, and in stark contrast to the usual case of coherent driving of multi-level systems, the resulting spin dynamics are strongly dependent on the phases $\phi_i$ ($i\in\{1,2,3\}$) of the driving fields, through the gauge-invariant, global phase $\Phi = \phi_1 + \phi_3 - \phi_2$.
In the following, we will examine the case of resonant, symmetric driving, for which $\delta_1=\delta_2=0$ and $\Omega_i=\Omega, \forall i$. In this case, $\hat{\boldsymbol{H}}_0$ can be readily diagonalised with resulting dressed eigenstates and eigenenergies
\begin{eqnarray}
	\ket{\Psi_k}&=& \frac{1}{\sqrt{3}}\left( e^{ i(\Phi/3+2 k \varphi_0)}, 1, e^{ -i(\Phi/3-k \varphi_0)} \right)
	\label{equ:EigenvectorsRWA}\\
	E_k / \hbar &=& \Omega \cos{\left(\Phi/3-k\varphi_0\right) }
	\label{equ:EigenenergiesRWA}
\end{eqnarray}
with $k \in \{ -1,0,1 \}$ and $\varphi_0=2\pi/3$. 

To experimentally observe CCI dynamics and generate the required, time-varying strain field, we place a single NV centre in a mechanical resonator of eigenfrequency $\omega_3/2\pi=9.2075~$MHz, which we resonantly drive using a nearby piezo-electric transducer\,\cite{Barfuss2015}. 
The mechanical Rabi frequency $\Omega_3$ is  controlled by the amplitude of the piezo excitation. 
To achieve resonant strain driving, i.e. $\omega_3/2 \pi=\Delta_Z$, we apply a static magnetic field $B_{\rm NV}$ along the NV axis. 
The two microwave magnetic fields used to address the $\ket{0} \leftrightarrow \ket{\pm1}$ transitions at frequencies $\omega_{1,2}=2\pi D_0 \pm \omega_3/2$  are delivered to the NV centre using a homebuilt near-field microwave antenna, (see SOM\,\cite{SOM} for information on phase control and microwave field generation). Finally, a confocal microscope is used for optical initialisation and readout of the NV spin (Fig.\,\ref{fig:fig1}b).

\begin{figure*}
	\centering
		\includegraphics{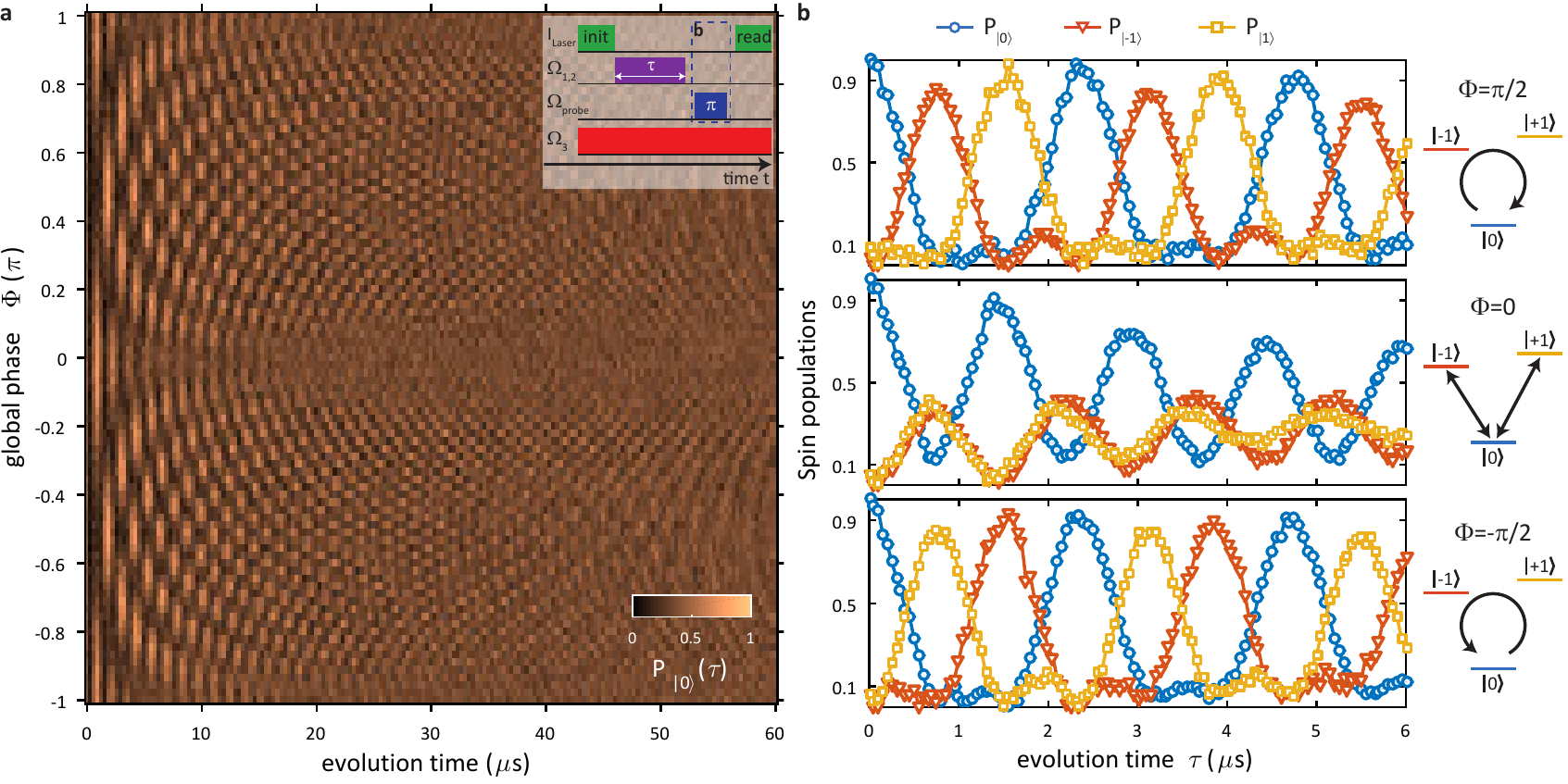}
	\caption{{\bf Time reversal symmetry breaking in closed-contour spin dynamics controlled by global phase $\Phi$}. a) Time evolution of $\ket{0}$ population, $P_{\ket{0}}$, as a function of global phase $\Phi$, after initialisation in $\ket{0}$. Closed-contour driving of the NV spin leads to periodic evolution of $P_{\ket{0}}$ due to quantum interference in the NV ground state. Period and decay times of the interference pattern depend strongly on $\Phi$. b) Linecuts of $P_{\ket{m_s}}(\tau)$ (with $m_s \in \lbrace -1, 0, +1 \rbrace$) for $\Phi=\pi/2, 0$ and $-\pi/2$ (top, middle and bottom panel, respectively). At phases $\Phi = \pm \pi/2$, population is shuffled clockwise or counterclockwise while for $\Phi = 0$ it alternates between $\ket{0}$ and an equal superposition of $\ket{\pm 1}$ (with some addmixture of $\ket{0}$).
	}
	\label{fig:fig2}
\end{figure*}

We study the NV spin dynamics under closed-contour driving by measuring the time evolution of the NV spin population for different values of $\Phi$, using the experimental sequence shown Fig.\,\ref{fig:fig2}a~(inset). 
For each value of $\Phi$, a green laser pulse initialises the NV spin in $\ket{\psi\left(\tau=0\right)}:=\ket{0} = \left(\ket{\Psi_{-1}}+\ket{\Psi_0}+\ket{\Psi_1}\right)/\sqrt3$, after which we let the system evolve under the influence of the three driving fields for a variable evolution time $\tau$. Finally, we apply a green laser pulse to read out the final population in $\ket{0}$, $P_{\ket{0}}(\tau) = |\langle0\ket{\psi\left(\tau\right)}|^2$, where $\ket{\psi\left(\tau\right)} = e^{-i \hat{H_0} \tau / \hbar} \ket{\psi\left(0\right)}$. 
The resulting data (Fig.\,\ref{fig:fig2}a), here for $\Omega/2\pi = 500~$kHz, show oscillations of $P_{\ket{0}}$ in time, with a marked $\pi$-periodic dependence of the population dynamics on $\Phi$.

To obtain a complete picture of the resulting spin-dynamics, we additionally monitor the populations $P_{\ket{\pm 1}}$ of spin-states $\ket{\pm 1}$ for $\Phi=0$ and $\pm\pi/2$ (Fig.\,\ref{fig:fig2}b) by applying a microwave $\pi$-pulse resonant with the $\ket{0}\rightarrow\ket{+1}$ or $\ket{0}\rightarrow\ket{-1}$ transition at the end of the evolution time $\tau$ ($\Omega_{\rm probe}$, dashed box in the inset of Fig.\,\ref{fig:fig2}a). 
The resulting spin dynamics show that at $\Phi = \pm \pi/2$ the spin exhibits time-inversion symmetry breaking circulation (Fig.\,\ref{fig:fig2}b, right) of population between the three states $\ket{0}$, $\ket{+1}$ and $\ket{-1}$\,\cite{Roushan2017}, with a period $T_{\pm \pi/2}=4\pi/\sqrt{3}\Omega$.
In perfect analogy to chiral currents of electrons hopping on a plaquette with three sites, threaded by a synthetic magnetic flux $\Phi$, this circulation of population demonstrates the tunable synthetic gauge field\,\cite{Roushan2017} created by our CCI driving scheme.
Conversely, for $\Phi = 0$ the spin level population oscillates between $\ket{0}$ and an equal superposition of $\ket{\pm1}$ in a ``V-shaped" trajectory (see Fig.\,\ref{fig:fig2}, middle) at a period $T_0 = 4 \pi /3 \Omega$. 
This shortening of $T_0$ compared to $T_{\pm \pi/2}$ is consistent with the different trajectories (Fig.\,\ref{fig:fig2}b, right) the spin populations undergo. 
To further support that Hamiltonian $\hat{\boldsymbol{H}}_0$ provides an accurate description of our system, we calculate the population dynamics and find excellent agreement with data (see SOM\,\cite{SOM} for details of the simulation and comparison with experiment).

\begin{figure}
	\centering
		\includegraphics{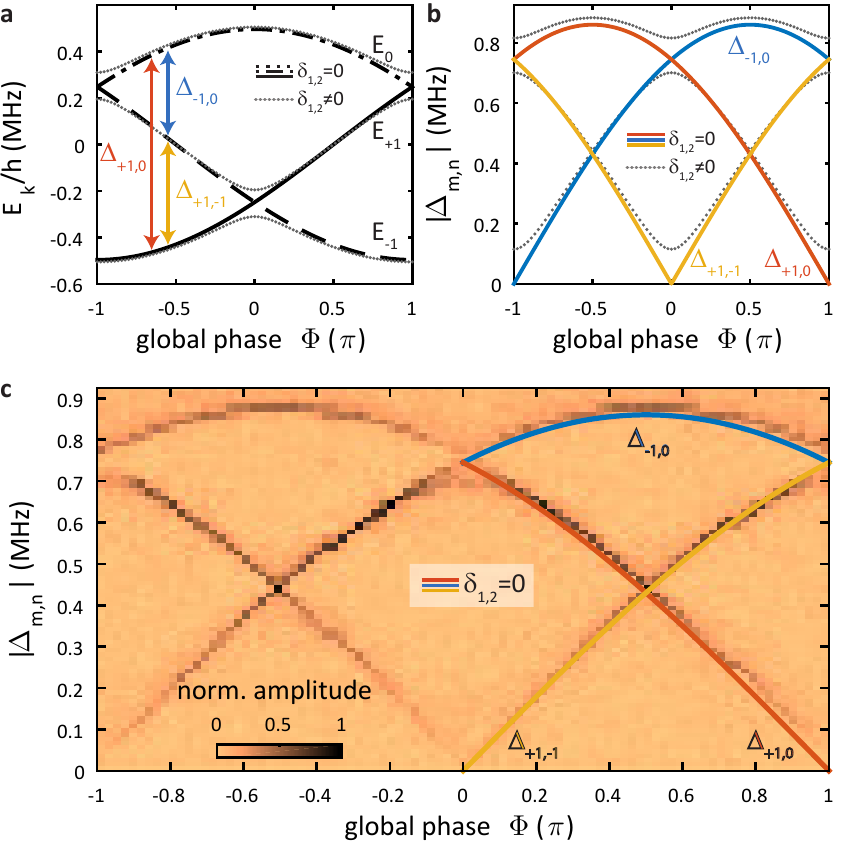}
 	\caption{{\bf Spectrum of the driven NV spin under closed-contour driving}. a) Calculated eigenenergies $E_{k}$ of the driven spin for $\Omega/2\pi=500~$MHz, as a function of $\Phi$ for detuning $\delta_{1,2}=0$ (black lines) and $\delta_{1,2}/2\pi=\pm50~$kHz (dotted lines). b) Transition frequencies $\lvert \Delta_{m,n} \rvert$ as a function of $\Phi$ for $\delta_{1,2}=0$ (blue, orange and red lines) and $\delta_{1,2}/2\pi=\pm50~$kHz (dotted lines). c) Discrete Fourier transform of the data shown in Fig.\,\ref{fig:fig2}a, as a function of $\Phi$. The spectral components observed agree well with the calculated values of  $\lvert \Delta_{m,n} \rvert$; discrepancies around $\Phi = 0,\pm\pi$ arise from environmental magnetic field fluctuations (see text). The observed Fourier amplitude (contrast) is inversely proportional to linewidth and therefore gives an indication of the decay time for each spectral component.}
	\label{fig:fig3}
\end{figure}

In addition to the spin dynamics under CCI, our experiment also allows us to directly access the eigenenergies $E_k$ of the driven three-level system (see Eq.\,(\ref{equ:EigenenergiesRWA}) and black lines in Fig.\,\ref{fig:fig3}a).
After initialisation into $\ket{0}=\left(\ket{\Psi_{-1}}+\ket{\Psi_0}+\ket{\Psi_1}\right)/\sqrt3$, each component $\ket{\Psi_k}$ acquires a dynamical phase $E_k \tau /\hbar$, which governs the time evolution of the NV spin.
The population $P_{\ket{0}}\!\!\left(\tau\right)$ therefore shows spectral components at frequencies $\Delta_{m,n} = \left( E_m-E_n \right) /h$ with $m\neq n \in \{-1,0,1\}$ (Fig.\,\ref{fig:fig3}a and b). 
A Fourier transformation of $P_{\ket{0}}(\tau)$ (Fig.\,\ref{fig:fig3}c) thus reveals $\Delta_{m,n}$ and thereby the eigenenergies of the driven NV spin, which for most values of $\Phi$ are in excellent agreement with the predictions based on $\hat{\boldsymbol{H}}_0$ (colored lines in Fig.\,\ref{fig:fig3}c).
Around $\Phi = 0$ and $\pm\pi$, we find anti-crossings instead of the expected frequency crossings in the spectrum; an observation we assign to environmental fluctuations and slow drifts.
Indeed, the resulting, non-resonant or asymmetric drive lifts the degeneracies of the dressed states and explains our observation (Fig.\,\ref{fig:fig3}a and b). 
Taking these effects into account, we conducted numerical modelling of our experiment and found good qualitative agreement with our observed spectra (see SOM).

\begin{figure}[t!]
	\centering
		\includegraphics{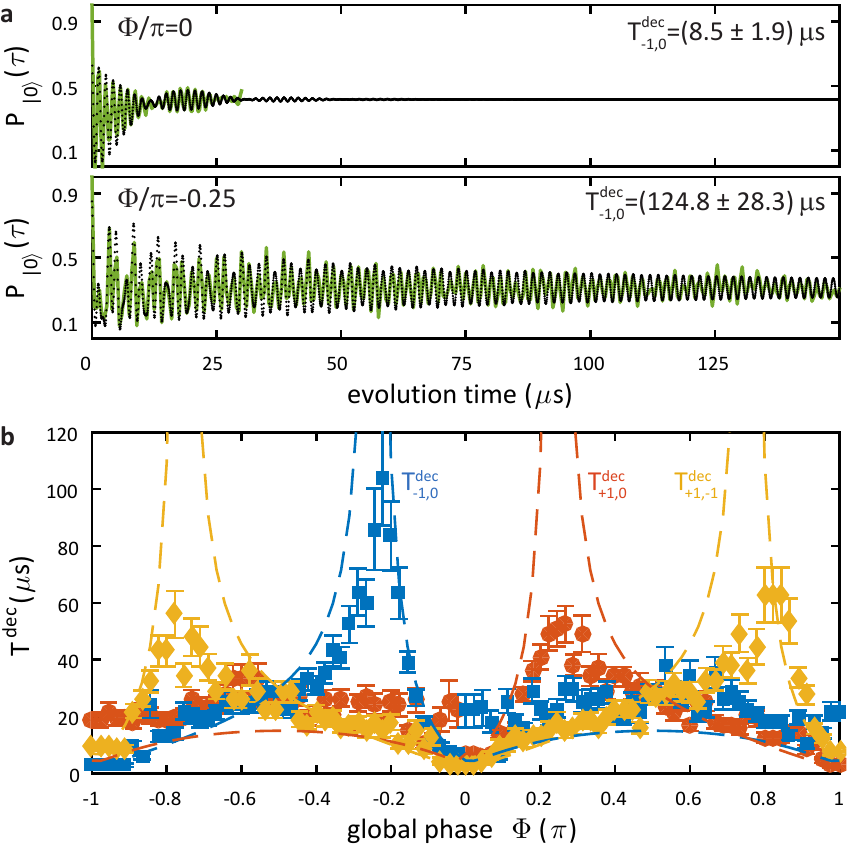}
 	\caption{{\bf Phase-controlled coherence protection}.
a) Spin-oscillations under closed-contour driving for $\Phi = 0$ and $\Phi = - \pi/4$, revealing strongly phase-dependent decay times $T^\mathrm{dec}_{m,n}(\Phi)$. 
A fit of exponentially damped harmonics (see text) yields $T^\mathrm{dec}_{-1,0}(\Phi = 0) = 8.5 \pm 1.9~\mu$s, and $T^\mathrm{dec}_{-1,0}(\Phi = - \pi/4) = 124.8 \pm 28.3~\mu$s for the most long-lived spectral components.
b) Systematic measurement of decay times as a function of $\Phi$, showing minima of $T^\mathrm{dec}_{m,n}(\Phi)$ at $\Phi/\pi = \pm1,0$ and pronounced maxima at $\Phi \approx \pm n \pi/4$, $n \in {1,3}$. Dashed lines are the results of a second-order perturbative calculation of $T^\mathrm{dec}_{m,n}(\Phi)$ (see text). 
Note that data in a) and b) originate from separate measurement runs and therefore result in slight differences in decay times. 
}
	\label{fig:fig4}
\end{figure}

The effect of environmental fluctuations is already visible in the phase dependent interference patterns in Fig.\,\ref{fig:fig2}a, where the resulting quantum-beats decay fastest for phase-values close to $\Phi = 0$ and $\pm\pi$ -- an indication that at these phase values, the dressed states $\ket{\Psi_k}$ are most vulnerable to environmental fluctuations, but protected from them at other values of $\Phi$. 
Fig.\,\ref{fig:fig4}a shows linecuts taken at $\Phi = 0$ (top panel) and $\Phi = -\pi/4$ (bottom panel), which evidence a dramatic change of the dressed state coherence time from $T^\mathrm{dec}_{-1,0} = \left(8.5\pm1.9\right)\,\mu s$ at $\Phi=0$ to $T^\mathrm{dec}_{-1,0} = \left(124.8\pm28.3\right)\,\mu s$ at  $\Phi = - \pi/4$.
To systematically quantify this $\Phi$-dependent dephasing, we fit a sum of three exponentially decaying sinusoids to the time-traces in Fig.\,\ref{fig:fig2}a and extract decay times $T^\mathrm{dec}_{m,n}$ for each frequency component $\Delta_{m,n}$.
The resulting dependence of $T^\mathrm{dec}_{m,n}$ on $\Phi$ is shown in Fig.\,\ref{fig:fig4}b, and exhibits pronounced maxima of Rabi decay times at  $\Phi \approx \pm n \pi/4$, $n \in \lbrace 1,3 \rbrace$\,\cite{SOM}.

To understand the phase-dependance of the dressed-state dephasing times, we conducted extensive numerical modelling together with perturbative, analytical calculations of $T^\mathrm{dec}_{m,n}(\Phi)$ (see SOM). 
Our second order perturbative calculations account for magnetic field fluctuations with Ornstein-Uhlenbeck statistics, together with a random field that was held static over each experimental run.
The result (dashed lines in Fig.\,\ref{fig:fig4}b) reveals that for each of the values $\Phi\approx\pm\pi/4$ and $\pm3\pi/4$, 
two dressed states exist whose energies show the same perturbative response to magnetic field fluctuations and thus form a coherence-protected subspace in the dressed state manifold, in which $T^\mathrm{dec}_{m,n}(\Phi)$ approaches the spin relaxation time. 
We assign the significantly reduced, measured value $T^\mathrm{dec}_{-1,0}(-\pi/4)\approx105 \, \mu$s to driving field fluctuations -- a hypothesis which we could quantitatively support with our numerical modelling (see SOM). 
Our data also show that the four local maxima of $T^\mathrm{dec}_{m,n}$ vary significantly in magnitude. 
We attribute this variation to slow experimental drifts of the zero field splitting parameter $D_0$ due to temperature variations\,\cite{Acosta2010} in our experiment. 
Taking these drifts into account in our model yields excellent agreement between simulation and experiment for realistic temperature variations of $\pm 1.3~$K (see SOM).

Our results establish the driving field phase under CCI driving as a novel control parameter for coherent manipulation and dynamical decoupling of single spins. 
They indicate that further experimental improvements would readily yield coherence protected dressed states with inhomogeneous dephasing times approaching the $T_1$-limit.
Such dressed states have recently been establishes as powerful resources for quantum sensing of GHz fields\,\cite{Joas2017,Stark2017}. The efficient tunability and coherence protection we demonstrate here for dressed states offer highly interesting avenues for enhanced sensitivities and phase-tuning of the sensing-frequencies for such sensing schemes.
In addition, our results yield an attractive platform to test, and ultimately implement proposed schemes for phase dependent CPT\,\cite{Kosachiov1992, Kosachiov1991, Windholz2001}, since the NV centre would readily allow for promoting CCI to the optical domain, where established optical $\Lambda$-transitions\,\cite{Zhou2017} could be combined with strain driving in a phase-coherent way.
Lastly, we note the strong analogy between the non-reciprocal spin dynamics under CCI driving we demonstrated and recent realisations of synthetic gauge fields in optomechanical systems\,\cite{Fang2017}.
Pursuing this analogy using ensembles of NV centres with engineered dissipation offers interesting avenues for realising on-chip, non-reciprocal microwave elements, such as microwave circulators or directional amplifiers\,\cite{Fang2017}.

\begin{acknowledgments}
We thank A. Retzker, N. Aharon, A. Nunnenkamp and H. Ribeiro for fruitful discussions and valuable input. We gratefully acknowledge financial support through the NCCR QSIT, a competence centre funded by the Swiss NSF, through the Swiss Nanoscience Institute, by the EU FP7 project DIADEMS (grant \#611143)
and through SNF Project Grant 169321.
\end{acknowledgments}

%

\pagebreak
\newcommand{\beginsupplement}{%
	\setcounter{table}{0}
	\renewcommand{\thetable}{S\arabic{table}}%
	\setcounter{figure}{0}
	\renewcommand{\thefigure}{S\arabic{figure}}%
}
\beginsupplement

\onecolumngrid
\section{Supplementary Information for \\
	"Non-reciprocal coherent dynamics of a single spin under closed-contour interaction"}

In the first part of the supplementary material, we give more background about the Hamiltonian describing our system, including some discussion of the population dynamics for different values of $\Phi$. We then simulate the effects of noise on the population dynamics and compare the results, including the phase-dependent dressed state coherence times, to the experimental data. In the second part, we discuss the perturbation terms for dressed state energies to first and second order in magnetic field noise along the NV axis. We find an analytical expression for the induced energy fluctuations as a function of $\Phi$, which we compare to the $\Phi$-dependence of the coherence times. Finally, in the third and last part we give additional experimental details on how we generate and control the microwave fields.

\section{Phase-dependent spin dynamics}

\label{sec:App_QuantumBeat}
\subsection{Closed-contour Hamiltonian and time evolution}
\label{subsec:App_QuantumBeat_DerivationHamiltonian}

In this section we derive the Hamiltonian (1) of the main text.
The NV's S=1 ground state, driven by two MW fields and one strain field, is described by the Hamiltonian
\begin{equation}
\hat{\boldsymbol{H}}_\mathrm{0}^\mathrm{lf}/\hbar =
\begin{pmatrix} 
\omega_{|-1\rangle} & \Omega_1 \cos\left( \omega_1 t + \phi_1 \right) & \Omega_3 \cos\left(\omega_3 t + \phi_3\right) \\
\Omega_1 \cos\left(\omega_1 t + \phi_1\right) &  \omega_{|0\rangle} & \Omega_2 \cos\left(\omega_2 t + \phi_2\right) \\ 
\Omega_3 \cos\left(\omega_3 t + \phi_3\right) & \Omega_2 \cos\left(\omega_2 t + \phi_2\right) & \omega_{|+1\rangle}
\end{pmatrix}
\label{eq:App_QuantumBeat_DerivationHamiltonian_Labframe}
\end{equation}
written in the $\{|-1\rangle,|0\rangle,|+1\rangle\}$ basis and expressed in the lab frame.
$\Omega_i$, $\omega_i$, and $\phi_i$ with $i = \lbrace 1,2,3 \rbrace$ denote driving field amplitudes, frequencies, and phases, respectively. $\hbar \omega_{|i\rangle}$ with $i = 0,\pm1$ are the energies of the three spin sublevels.

We transform $\hat{\boldsymbol{H}}_\mathrm{0}^\mathrm{lf}$ into the interaction picture by performing the unitary transformation \cite{Buckle1986, Pegg1986}
\begin{equation}
\hat{\boldsymbol{H}}_\mathrm{0} = \hat{\boldsymbol{T}} \hat{\boldsymbol{H}}_\mathrm{0}^\mathrm{lf} \hat{\boldsymbol{T}}^{-1} + \mathrm{i}  \frac{\mathrm{d} \hat{\boldsymbol{T}}}{\mathrm{d} t} \hat{\boldsymbol{T}}^{-1}
\label{eq:App_QuantumBeat_DerivationHamiltonian_unitarytransformation}
\end{equation}
with the unitary rotation operator 
\begin{equation}
\hat{\boldsymbol{T}} = e^{ \mathrm{i} \left(\omega_1 t + \phi_1\right) |-1\rangle\langle-1|}
\cdot
e^{\mathrm{i}  \left(\omega_2 t + \phi_2\right)|+1\rangle\langle+1|}.
\label{eq:App_QuantumBeat_DerivationHamiltonian_unitaryoperator}
\end{equation}
By choosing $\hbar \omega_{|0 \rangle} = 0$ and neglecting fast rotating terms, we find
\begin{equation}
\hat{\boldsymbol{H}}_\mathrm{0} /\hbar = \frac{1}{2}
\begin{pmatrix} 
2( \omega_{|-1\rangle} - \omega_1) & \Omega_1 & \Omega_3 e^{\mathrm{i}  \left( \Phi + \Delta \, * \, t \right) } \\
\Omega_1  & 0 & \Omega_2 \\ 
\Omega_3 e^{-\mathrm{i}  \left( \Phi + \Delta \, * \, t \right) }   & \Omega_2 & 2( \omega_{|+1\rangle} - \omega_2)
\end{pmatrix}
\label{eq:App_QuantumBeat_DerivationHamiltonian_RWA}
\end{equation}
where the global phase $\Phi = \phi_1 + \phi_3 - \phi_2$ and $\Delta = \omega_1 + \omega_3 - \omega_2$.
It becomes clear that $\hat{\boldsymbol{H}}_\mathrm{0}$ is time-independent only for $\Delta = 0$.
Consequently, for $\Delta=0$, equal driving strengths $\Omega_1 = \Omega_2 = \Omega_3 = \Omega$, and detunings $\delta_{1} = \omega_{|-1\rangle} - \omega_1$ and $\delta_{2} = \omega_{|+1\rangle} - \omega_2$, we obtain
\begin{equation}
\hat{\boldsymbol{H}}_\mathrm{0} /\hbar = \frac{1}{2}
\begin{pmatrix} 
2 \delta_1 & \Omega& \Omega e^{\mathrm{i}  \Phi } \\
\Omega  & 0 & \Omega \\ 
\Omega e^{-\mathrm{i} \Phi }   & \Omega & 2 \delta_2
\end{pmatrix},
\label{eq:App_QuantumBeat_DerivationHamiltonian_RWAasintext}
\end{equation}

\noindent as given in the main text.
The corresponding eigenstates and eigenenergies for resonant ($\delta_1 = \delta_2 = 0$) driving are then found to be
\begin{eqnarray}
|\Psi_k\rangle &=& \frac{1}{\sqrt{3}} \left( e^{\mathrm{i}  \frac{ \Phi + k 4\pi}{3} }, 1, e^{ -\mathrm{i}  \frac{\Phi - k2\pi}{3} }\right)
\label{App_QuantumBeat_DerivationHamiltonian_Eigenvectors}\\
E_k/\hbar &=& \Omega \cos{\left( \frac{\Phi-k2\pi}{3} \right) }
\label{App_QuantumBeat_DerivationHamiltonian_Eigenenergies}
\end{eqnarray}
where $k \in \{ -1,0,1 \}$. 

In the main text, we study the time evolution of the population $P_{|0\rangle}(\tau)$ in spin sublevel $|0\rangle$.
In the corresponding experiments we initialize the NV in $|0\rangle$ with a green laser pulse.
The following abrupt onset of the three driving fields creates the closed-contour interaction, and we express 
\begin{equation}
|0\rangle = |\Psi\left(\tau=0\right)\rangle = \left(|\Psi_{-1}\rangle + |\Psi_{0}\rangle + |\Psi_{+1}\rangle\right)/\sqrt3
\label{eq:App_QuantumBeat_DerivationTimeEvolution_ms0}
\end{equation}
as a linear combination of the system's eigenstates $|\Psi_k \rangle$.
After an evolution time $\tau$ the system is in the state
\begin{equation}
\begin{split}
|\Psi\left(\tau\right)\rangle & =  e^{-\mathrm{i}  \hat{\boldsymbol{H}}_\mathrm{0} \tau/\hbar}|\Psi\left(\tau = 0\right)\rangle \\
& = \left(e^{-\mathrm{i}  E_0 \tau/\hbar}|\Psi_0\rangle +  e^{-\mathrm{i}  E_{+1} \tau/\hbar}|\Psi_{+1}\rangle  +  e^{-\mathrm{i}  E_{-1} \tau/\hbar}|\Psi_{-1}\rangle\right)/\sqrt3 
\end{split}
\label{eq:App_QuantumBeat_DerviationTimeEvolution_psit}
\end{equation}
where $\hat{\boldsymbol{U}} = e^{-\mathrm{i} \hat{\boldsymbol{H}}_0 \tau/\hbar}$ is the unitary time evolution operator.
The final green laser pulse  yields a measurement of the resulting population in $|0\rangle$ and we determine
\begin{equation}
\begin{split}
P_{|0\rangle}\left(\tau\right) &= |\langle0|\Psi\left(\tau\right)\rangle|^2  \\
&=\frac{1}{3} + \frac{2}{9} 
\left[ 
\cos \left( 2\pi\Delta_{-1,0}\tau \right) + 
\cos \left( 2\pi\Delta_{+1,0}\tau \right) + 
\cos \left( 2\pi\Delta_{+1,-1}\tau \right) 
\right]
\end{split}
\label{eq:App_QuantumBeat_DerviationTimeEvolution_Pms0}
\end{equation} 
with $\Delta_{i,j} = (E_{i}-E_{j})/h$. $P_{|0\rangle}\left(\tau\right)$ therefore oscillates due to the presence of three frequency components $\Delta_{i,j}$ that correspond to the differences of eigenenergies $E_k$. At $\Phi = 0,\pm\pi$\,,
\begin{equation}
P_{|0\rangle}^{0,\pm\pi}\left(\tau\right) = \frac{1}{9}\left[ 5 + 4 \cos\left(\frac{3\Omega}{2} \tau \right) \right] 
\label{eq:App_QuantumBeat_DerviationTimeEvolution_Pms0_Phi0}
\end{equation} 
and spin population oscillates between $|0\rangle$ and a state composed of an equal superposition of $|\pm1\rangle$ with a small admixture of $\lvert 0 \rangle$ at a period $T_{0,\pm\pi} = 4\pi/3\Omega$ (see Fig.~2a of the main text). Eq.\,\eqref{eq:App_QuantumBeat_DerviationTimeEvolution_Pms0_Phi0} also demonstrates that $P_{|0\rangle}^{0,\pm\pi}\left(\tau\right) \neq 0$ for any time $\tau$. At $\Phi = \pm\pi/2$, however,
\begin{equation}
P_{|0\rangle}^{\pm\pi/2}\left(\tau\right) = \frac{1}{9}\left[ 3 + 4 \cos\left(\frac{\sqrt{3}\Omega}{2}\tau\right) + 2 \cos\left(\sqrt{3}\Omega\tau\right) \right] 
\end{equation} 
and $P_{|0\rangle}^{\pm\pi/2}\left(\tau\right)$ is maximized every $T_{\pm\pi/2} = 4\pi/\sqrt{3}\Omega$.

To extract the frequency components and decay times from the data, we fit our measured values of $P_{|0\rangle}(\tau)$ with a sum of three exponentially decaying sinusoids:

\begin{equation}
P_{|0\rangle}(\tau) = C + \sum_{n = 1}^3 A_n \, \cos{ \left( \omega_n \tau + \phi_n \right) } \, e^{- \tau / T_n}.
\end{equation}

\noindent Doing this for each value of $\Phi$ yields the phase-dependence of Rabi decay times $T^\mathrm{dec}_{i,j}$ (see Fig.~\ref{fig:App_QuantumBeat_NoiseSources_Overview_Comparions_FFTTdecay}).

\subsection{Analysis of environmental noise sources}
\label{sec:App_QuantumBeat_NoiseSources_Overview}

\subsubsection{Overview of existing noise sources}
\begin{figure}
	\centering
	\includegraphics[]{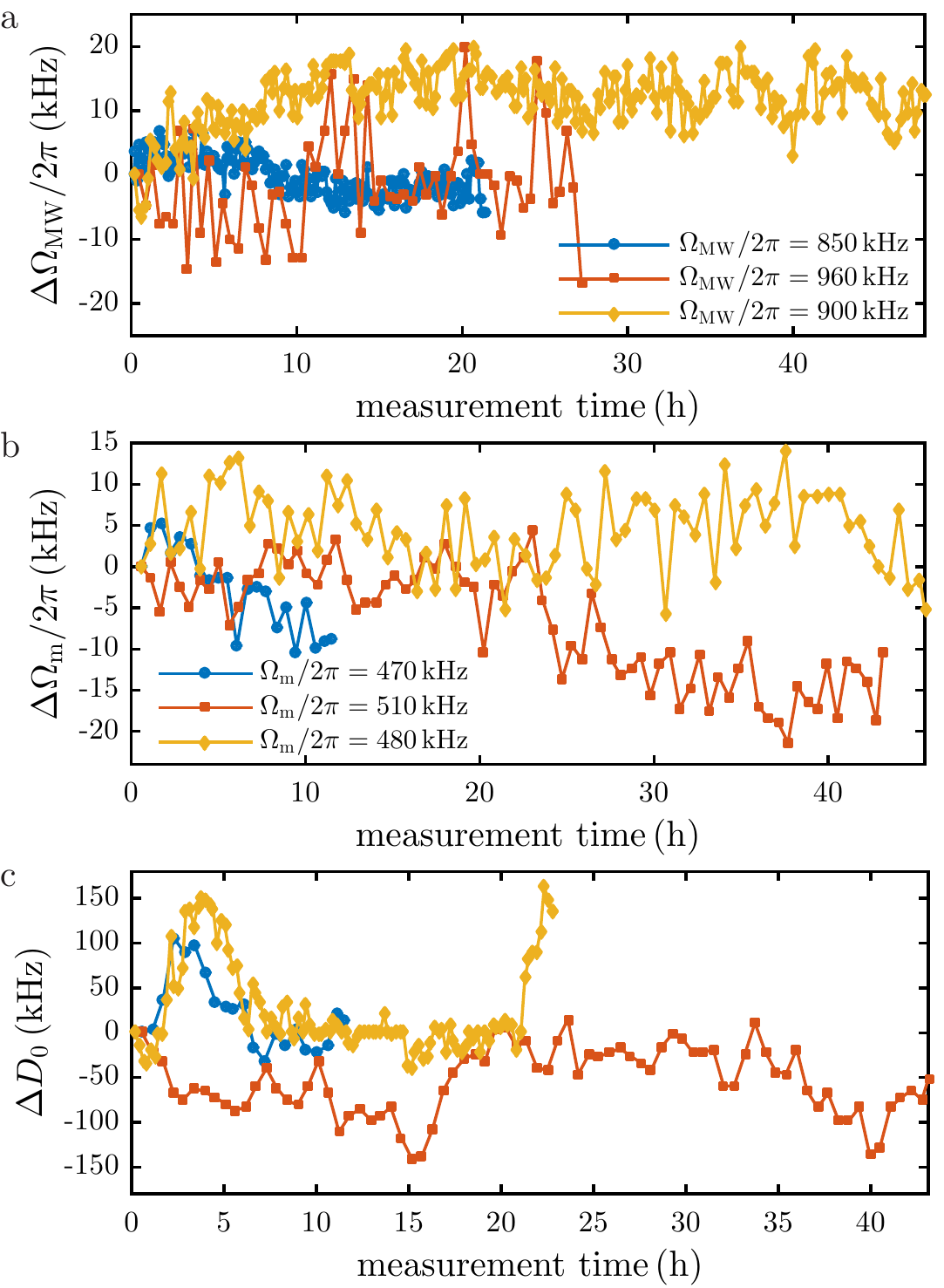}
	\caption{Existing fluctuations in our closed-contour interaction scheme.
		a) Typical fluctuations in our MW driving fields, measured for three different driving strengths $\Omega_\mathrm{MW}/2\pi = \{850, 900, 960 \}$\,kHz (blue, yellow and red curve, respectively) by performing Rabi oscillation measurements on the $|0\rangle \leftrightarrow |-1\rangle$ transition.
		b) Typical fluctuations in our AC strain field, measured for three different driving strengths $\Omega_\mathrm{m}/2\pi = \{470, 480, 510 \}$\,kHz (blue, yellow and red curve, respectively) via the Autler-Townes splitting of the $|+1,+1\rangle$ hyperfine level \cite{Barfuss2015}. 
		c) Fluctuations of the zero-field splitting $D_0$ for different measurement times for a single NV.
	}
	\label{fig:App_QuantumBeat_NoiseSources_Overview}
\end{figure}

The observed spin dynamics under closed-contour interaction and the dependence of Rabi decay times $T^\mathrm{dec}_{i,j}$ on driving phase $\Phi$ are significantly affected by the presence of several noise sources in our experiment. Most importantly, our measurements are influenced by fluctuations in
\begin{itemize}
	\item the environmental magnetic field, caused for example by nearby nuclear $^{14}$N or $^{13}$C spins and characterized by the NV's inhomogeneous coherence time $T_2^*$
	\item the amplitudes of our MW driving fields $\Omega_{1,2}$ (in the following referred to as $\Omega_\mathrm{MW}$), caused by technical noise in the MW circuit
	\item the AC strain driving strength $\Omega_3$ (labeled with $\Omega_\mathrm{m}$), originating from technical noise in the piezo driving signal
	\item the zero-field splitting $D_0$, caused by variations in temperature or environmental strain or electric fields. 
\end{itemize}
Other noise sources, for example frequency noise of the driving fields, are neglected in the following as we could not find any experimental evidence for their relevance.

To characterize the existing fluctuations in driving field amplitudes and zero-field splitting we performed long-time measurements, using the NV centre as a probe (Fig.\,\ref{fig:App_QuantumBeat_NoiseSources_Overview}).
Low frequency drifts in $\Omega_\mathrm{MW}$ were analyzed by performing Rabi oscillation measurements on a single hyperfine transition.
Similarly, slow fluctuations in $\Omega_\mathrm{m}$ and $D_0$ were investigated via the amplitude of strain-induced Autler-Townes splittings and ESR transition frequencies, respectively (see Barfuss {\textit{et al.}}~\cite{Barfuss2015}). 
We found relative driving amplitude fluctuations of $\sigma_\mathrm{MW}/\Omega_\mathrm{MW} = 0.5$\,\% and $\sigma_\mathrm{m}/\Omega_\mathrm{m} = 0.7$\,\% within a bandwidth of $\approx 2$\,mHz, with $\sigma$ being the corresponding standard deviations, assuming Gaussian distributions. Additionally, these fluctuations are accompanied by even slower amplitude drifts of $\delta_{\Omega_\mathrm{m}}^\prime/\Omega_\mathrm{m} \approx \pm1$\,\% and $\delta_{\Omega_\mathrm{MW}}^\prime/\Omega_\mathrm{MW} \approx \pm1.5$\,\% for strain and MW driving, respectively, which happen on timescales of several hours.
In contrast, changes in the zero-field splitting $\delta_{D_0}/2\pi$ are characterized solely by slow drifts, with drift amplitudes varying between a few and several tens of kHz. 
The data presented in the main text was taken on an NV with coherence time $T_2^* = (2.1 \pm 0.1)$\,$\mu$s, which we determined through Ramsey interference. Magnetic noise is thus characterized by a Gaussian distribution with width $\sigma_{T_2^*}/2\pi = 1/\sqrt{2}\pi T_2^* = 107$\,kHz \cite{Jamonneau2016} and is the strongest noise source in our experimental setting. 

In the experiments presented in the main text, we worked with $\Omega_{\mathrm{MW}}/2\pi = \Omega_\mathrm{m}/2\pi = 500$\,kHz.
Absolute driving field fluctuations $\delta_{\Omega_\mathrm{MW}}$ and $\delta_{\Omega_\mathrm{m}}$, characterized by standard deviations $\sigma_\mathrm{MW}/2\pi = 2.5$\,kHz and $\sigma_\mathrm{m}/2\pi = 3.5$\,kHz, as well as slow amplitude drifts $\delta_{\Omega_\mathrm{m}}^\prime/2\pi \approx \pm 5$\,kHz and $\delta_{\Omega_\mathrm{MW}}^\prime/2\pi \approx \pm 7.5$\,kHz, accompany fluctuations in the Zeeman splitting ($\sigma_{T_2^*}/2\pi = 107$\,kHz) and zero-field splitting ($\delta_{D_0}/2\pi \lesssim 100$\,kHz).
In our simulations, we neglect slow driving field amplitude fluctuations $\delta_{\Omega_\mathrm{MW}}^\prime$ and $\delta_{\Omega_\mathrm{m}}^\prime$ when modeling the influence of existing noise sources for reasons of simplicity, but consider faster driving amplitude fluctuations $\delta_{\Omega_\mathrm{MW}}$ and $\delta_{\Omega_\mathrm{m}}$.

\subsubsection{Modeling existing noise sources}

To simulate the influence of noise on the observed spin dynamics we average over $N_\mathrm{avg}$ solutions of the time-dependent Schr\"odinger equation
\begin{equation}
i\hbar\frac{\mathrm{d}}{\mathrm{d}t}|\Psi(\tau)\rangle = \hat{\boldsymbol{H}}_0(\Phi)|\Psi(\tau)\rangle
\end{equation}
where each solution is obtained for a different $\hat{\boldsymbol{H}}_0$.
In particular, we set
\begin{subequations}
	\label{eq:App_QuantumBeat_NoiseSources_DrivingFieldFluctuations}
	\begin{align}
	\Omega_{1,2} & = \Omega + \delta_{\Omega_\mathrm{MW}} \\
	\Omega_{3} & = \Omega + \delta_{\Omega_\mathrm{m}}
	\end{align}
\end{subequations}
where $\Omega/2\pi = 500$\,kHz denotes the applied Rabi frequency and $\delta_{\Omega_{\mathrm{MW}}}$ and $\delta_{\Omega_{\mathrm{m}}}$ describe Gaussian fluctuations, taken from a normal distribution with zero mean and standard deviations $\sigma_\mathrm{MW}/2\pi = 2.5$\,kHz and $\sigma_\mathrm{m}/2\pi = 3.5$\,kHz, as determined above, respectively. 
Magnetic and zero-field splitting fluctuations are included by setting
\begin{subequations}
	\label{eq:App_QuantumBeat_NoiseSources_Detunings}
	\begin{align}
	\delta_1 & = \delta_{D_0} + \delta_{T_2^*}  \\
	\delta_2 & = \delta_{D_0} - \delta_{T_2^*}
	\end{align}
\end{subequations}
in $\hat{\boldsymbol{H}}_0$. Variations $\delta_{D_0}$ in $D_0$ appear as simultaneous shifts of detunings $\delta_{1,2}$ while magnetic fluctuations $\delta_{T_2^*}$ induce an opposite change. $\delta_{D_0}$ can in principle be modeled by a random walk approach, but for our simulations presented in the following we usually set it manually. $\delta_{T_2^*}$ is taken from a normal distribution with zero mean and standard deviation $\sigma_{T_2^*} /2\pi= 107$\,kHz (see above). 

When solving the time-dependent Schr\"odinger equation, we update $\delta_{T_2^*}$, $\delta_{\Omega_{\mathrm{MW}}}$ and $\delta_{\Omega_{\mathrm{m}}}$ every step along $\Phi$, i.e.\,$N_\Phi$ times per complete phase sweep. In contrast, $\delta_{D_0}$ is changed at a much smaller rate, because the zero-field splitting changes on timescales of several hours (see Fig.~\ref{fig:App_QuantumBeat_NoiseSources_IndSweeps_RMS_Jan2017} to justify this approach). Please refer to the provided plots of $\delta_{D_0}(\Phi)$ (see Fig.\,\ref{fig:App_QuantumBeat_NoiseSources_Overview_Comparions_InfluenceZFS}). 
The presented procedure is repeated $N_\mathrm{avg}$ times and averaging over all solutions yields mean values for $|\Psi(\tau)\rangle$ and $E_k$ with $k = 0,\pm1$. Note that this approach limits us to low frequency fluctuations with a bandwidth of $1/\tau_\mathrm{max} \approx 15$\,kHz, as the experimental environment is kept constant as long as the global phase $\Phi$ remains unchanged ($\tau_\mathrm{max} \approx 60$\,$\mu$s the maximum evolution time for which $|\Psi(\tau)\rangle$ is calculated). The excellent agreement between simulations and experiment (see Sec.\,\ref{subsubsec:App_QuantumBeat_NoiseSources_Overview_Comparison}) justify our approach.

\subsubsection{Comparing simulation and experiment}
\label{subsubsec:App_QuantumBeat_NoiseSources_Overview_Comparison}

\begin{figure}
	\centering
	\includegraphics[]{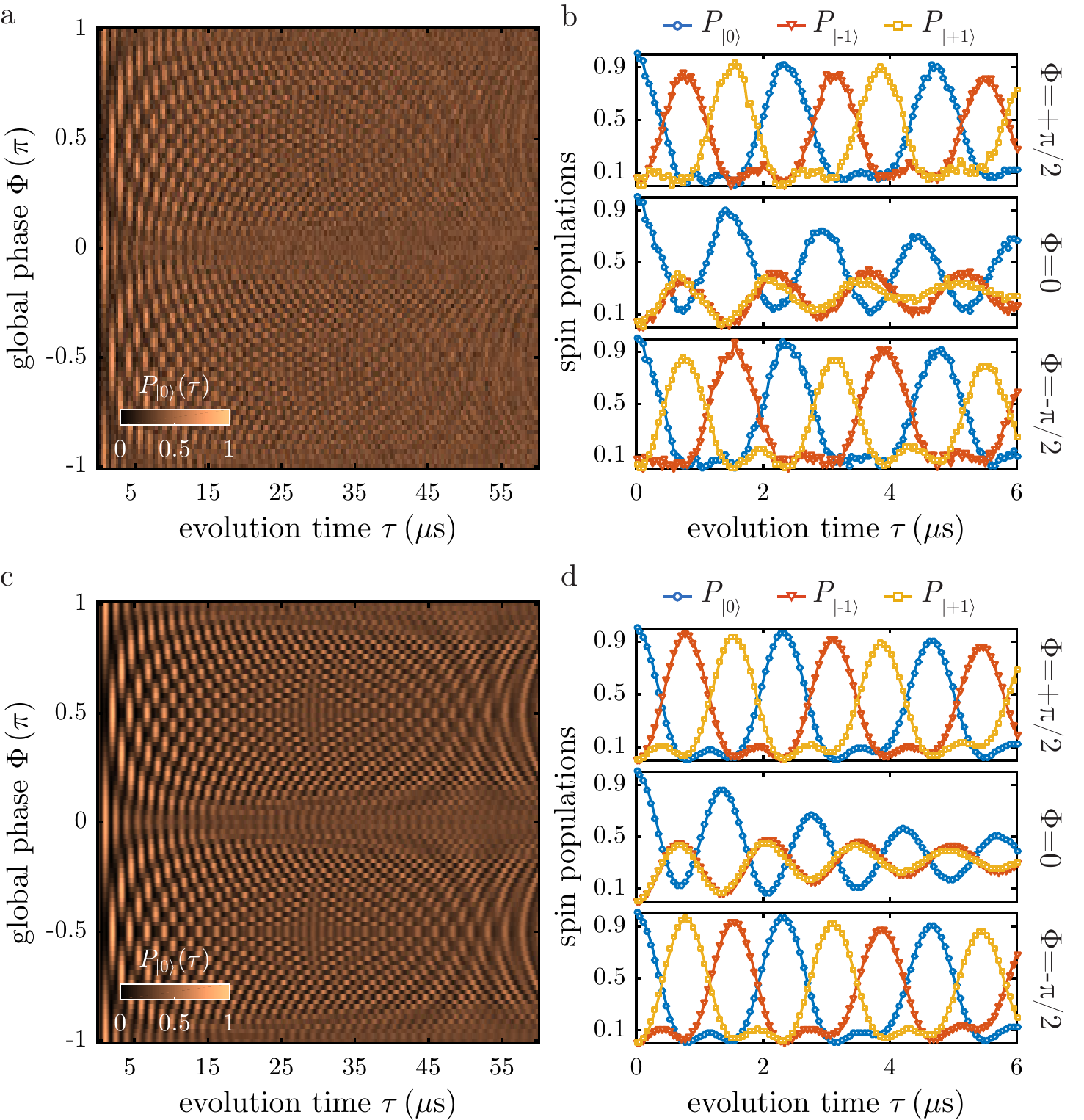}
	\caption{Closed-contour spin dynamics controlled by global phase $\Phi$ -- a comparison between experiment (panels a+b) and our model (panels c+d). In general, we observe excellent agreement between experimental data and simulations in the time domain.
		Furthermore, the experimentally obtained $P_{|\pm1\rangle}(\tau)$ in panel b are characterized by smaller oscillation amplitudes than predicted by our simulations. The disagreement is caused the fact that the strain field is still applied during the weak microwave swap-pulse used to readout the spin populations $P_{|\pm1\rangle}$. We were unable to quickly switch off the strain field because the finite Q factors of our mechanical resonators leads to a long response time.
		Finally, we want to point out that, due to the finite sampling rate of $\Phi$, the linecuts for $\Phi = 0$ in panel b have not been taken exactly at $\Phi = 0$, but slightly offset at $\Phi \approx +2^\circ$, causing the observed small mismatch in oscillation frequencies, weaker damping, and the non-degeneracy of $P_{|\pm1\rangle}$.}
	\label{fig:App_QuantumBeat_NoiseSources_Overview_Comparions_SpinDynamics}
\end{figure}

\begin{figure}
	\centering
	\includegraphics[]{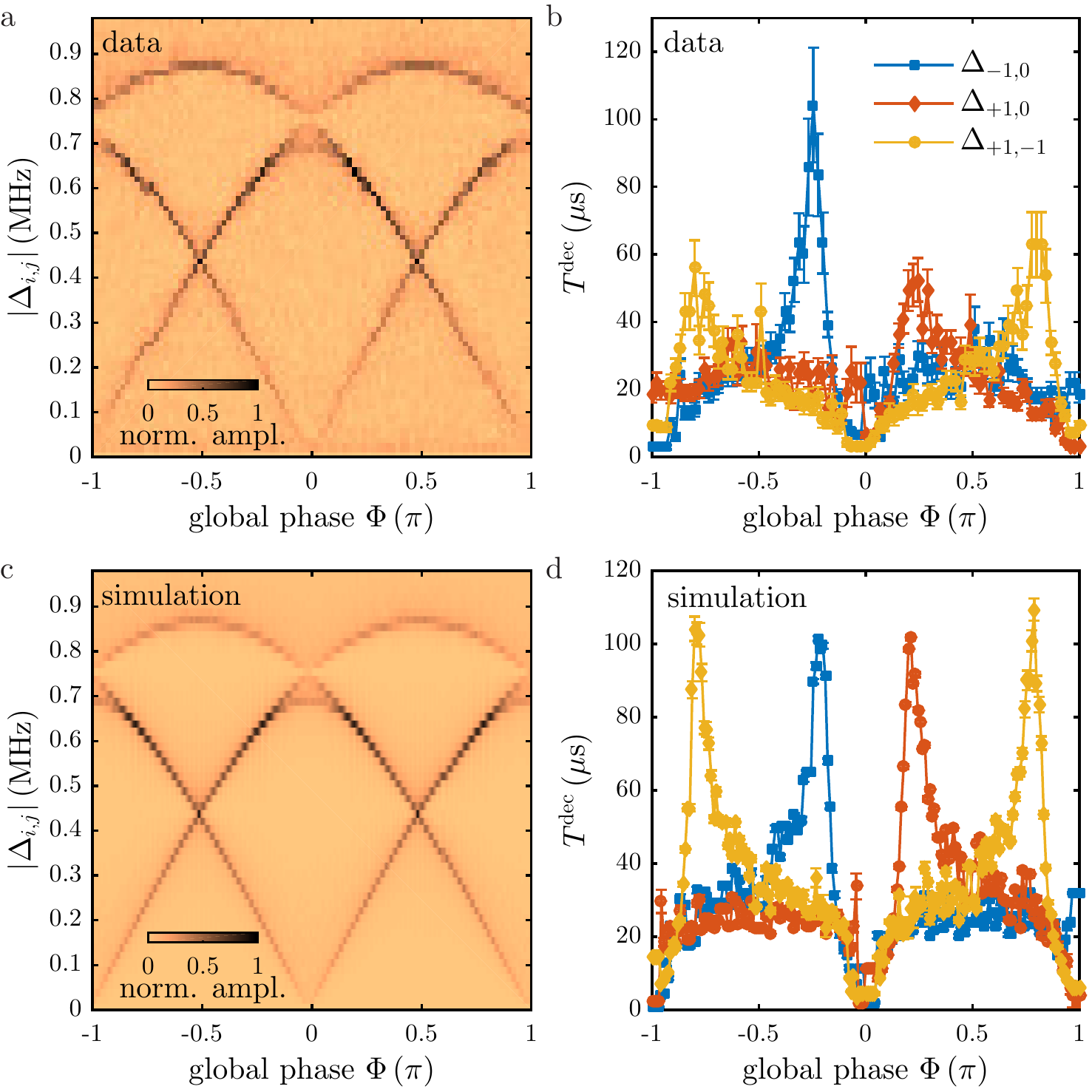}
	\caption{Spin precession frequencies and corresponding decay times under closed-contour interaction as measured experimentally (a+b) and obtained from simulations (c+d).
		Regarding the spin precession frequency spectra, we find great agreement between experiment and model. However, our data in panel a feature avoided crossings of varying gap sizes at $|\Delta_{i,j}| \approx 750\,$kHz for $\Phi / \Pi = 0,\pm 1$. In the simulated counterpart, (panel c) such gaps are identical.
		In addition, the Rabi decay time spectra in panels b and d are substantially different. While the modeled behavior of $T^\mathrm{dec}$ features four, equally long living frequency components at $\Phi \approx \pm 0.2 \pi, \pm 0.8 \pi$, the experimentally measured dependence of $T^\mathrm{dec}$ on $\Phi$ is strongly asymmetric. This difference is caused by slow fluctuations, most likely in the zero-field spitting $D_0$ (see text). Error bars denote $95$\,\% fit confidence intervals.}
	\label{fig:App_QuantumBeat_NoiseSources_Overview_Comparions_FFTTdecay}
\end{figure}

In Fig.\,\ref{fig:App_QuantumBeat_NoiseSources_Overview_Comparions_SpinDynamics} and Fig.\,\ref{fig:App_QuantumBeat_NoiseSources_Overview_Comparions_FFTTdecay}, we compare experimental data presented in the main text and our modeled results, which we obtained as described above (note that no fluctuations in the zero-field splitting $D_0$ are included unless stated otherwise).
In the time domain (Fig.\,\ref{fig:App_QuantumBeat_NoiseSources_Overview_Comparions_SpinDynamics}), we observe excellent agreement between experiment and model. Yet a few differences do exist.
First, the measured interference pattern of $P_{|0\rangle}(\tau)$ in Fig.\,\ref{fig:App_QuantumBeat_NoiseSources_Overview_Comparions_SpinDynamics}a is characterized by a slightly lower oscillation contrast compared to its simulated counterpart in Fig.\,\ref{fig:App_QuantumBeat_NoiseSources_Overview_Comparions_SpinDynamics}d. We assign this difference to the limited signal-to-noise ratio of our experimental data due to finite integration time. Also, slow fluctuations in driving fields and zero-field splitting have been neglected in our simulation.
Second, the experimentally obtained $P_{|\pm1\rangle}(\tau)$ in Fig.\,\ref{fig:App_QuantumBeat_NoiseSources_Overview_Comparions_SpinDynamics}b are characterized by slightly smaller oscillation amplitudes than predicted by our simulations. The simplified noise environment in our model is partly responsible for this discrepancy, as is the fact that the strain field is still applied during the weak microwave swap-pulse applied to read out spin populations $P_{\lvert \pm 1 \rangle}$. We are unable to quickly switch off the strain field due to the non-zero Q factors of our resonators, which leads to a long mechanical response time.

Finally, we want to point out that the linecuts for $\Phi = 0$ in Fig.\,\ref{fig:App_QuantumBeat_NoiseSources_Overview_Comparions_SpinDynamics}b have not been taken exactly at $\Phi = 0$, but slightly offset at $\Phi \approx +2^\circ$, due to finite sampling rate of $\Phi$. As a consequence, $P_{|\pm1\rangle}$ are not degenerate.

\begin{figure}
	\centering
	\includegraphics[]{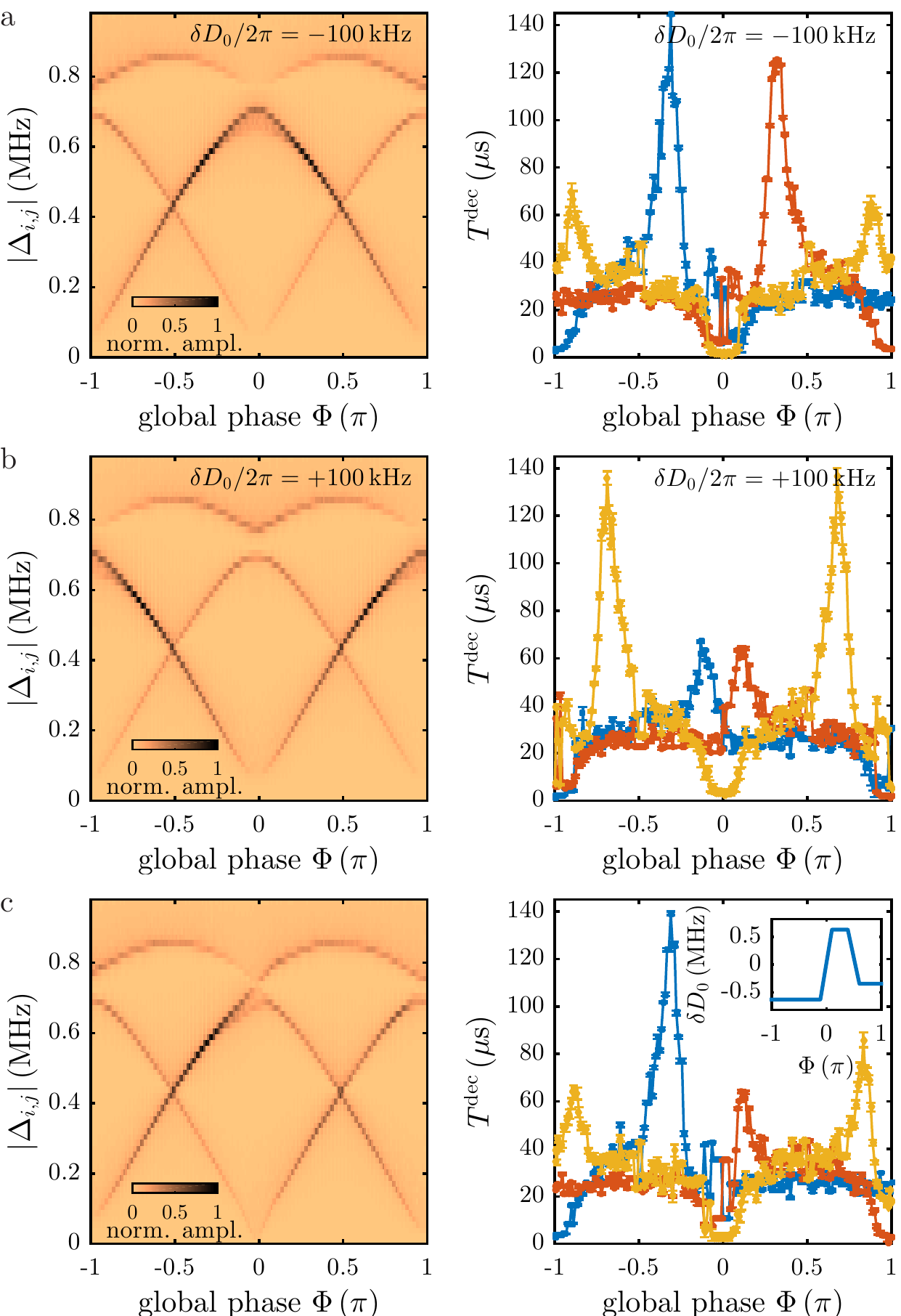}
	\caption{Influence of zero-field splitting variations on spin precession frequencies (left) and Rabi decay times (right). 
		a+b) A constant detuning $\delta D_0$ causes a strong asymmetry in the phase dependence of $T^\mathrm{dec}$. When $\delta D_0 > 0$, the $\Delta_{+1,-1}$ components at $\Phi \approx \pm 3\pi/4$ decay much slower than the $\Delta_{\pm1,0}$ components, located at $\Phi \approx \pm \pi/4$. For $\delta D_0 < 0$, this effect is reversed.
		c) Varying $\delta D_0$ with global phase $\Phi$ (see inset), allows reproducing the experimentally determined behavior from Fig.\,\ref{fig:App_QuantumBeat_NoiseSources_Overview_Comparions_FFTTdecay}b.}
	\label{fig:App_QuantumBeat_NoiseSources_Overview_Comparions_InfluenceZFS}
\end{figure}

Regarding the precession frequencies of the driven NV spin (Fig.\,\ref{fig:App_QuantumBeat_NoiseSources_Overview_Comparions_FFTTdecay}a+c), we again find remarkable agreement between experiment and model. However, our data in Fig.\,\ref{fig:App_QuantumBeat_NoiseSources_Overview_Comparions_FFTTdecay}a feature avoided crossings of varying gap sizes ($\sim 20$\,kHz at $\Phi = 0$; $\sim 100$\,kHz at $\Phi = \pm \pi$). In the simulated precession frequency spectrum (Fig.\,\ref{fig:App_QuantumBeat_NoiseSources_Overview_Comparions_FFTTdecay}c), however, all gaps are of equal size. As we will see later (see Fig.\,\ref{fig:App_QuantumBeat_NoiseSources_Overview_Comparions_InfluenceZFS}), we assign this mismatch to slow fluctuations in the zero-field splitting $D_0$.
In contrast to the excellent agreement in Fig.\,\ref{fig:App_QuantumBeat_NoiseSources_Overview_Comparions_FFTTdecay}a+c, the Rabi decay time spectra in Fig.\,\ref{fig:App_QuantumBeat_NoiseSources_Overview_Comparions_FFTTdecay}b+d are substantially different. The modeled phase dependence of $T^\mathrm{dec}$ features four, equally long living frequency components with maximum decay times of $T^\mathrm{dec} \approx 105$\,$\mu$s at $\Phi = \pm \pi/4, \pm 3\pi/4$. In our experimental data, $T^\mathrm{dec}$ is strongly asymmetric. Again, this difference is caused by slow fluctuations, most likely in the zero-field spitting $D_0$.

\begin{figure}[h!]
	\centering
	\includegraphics[]{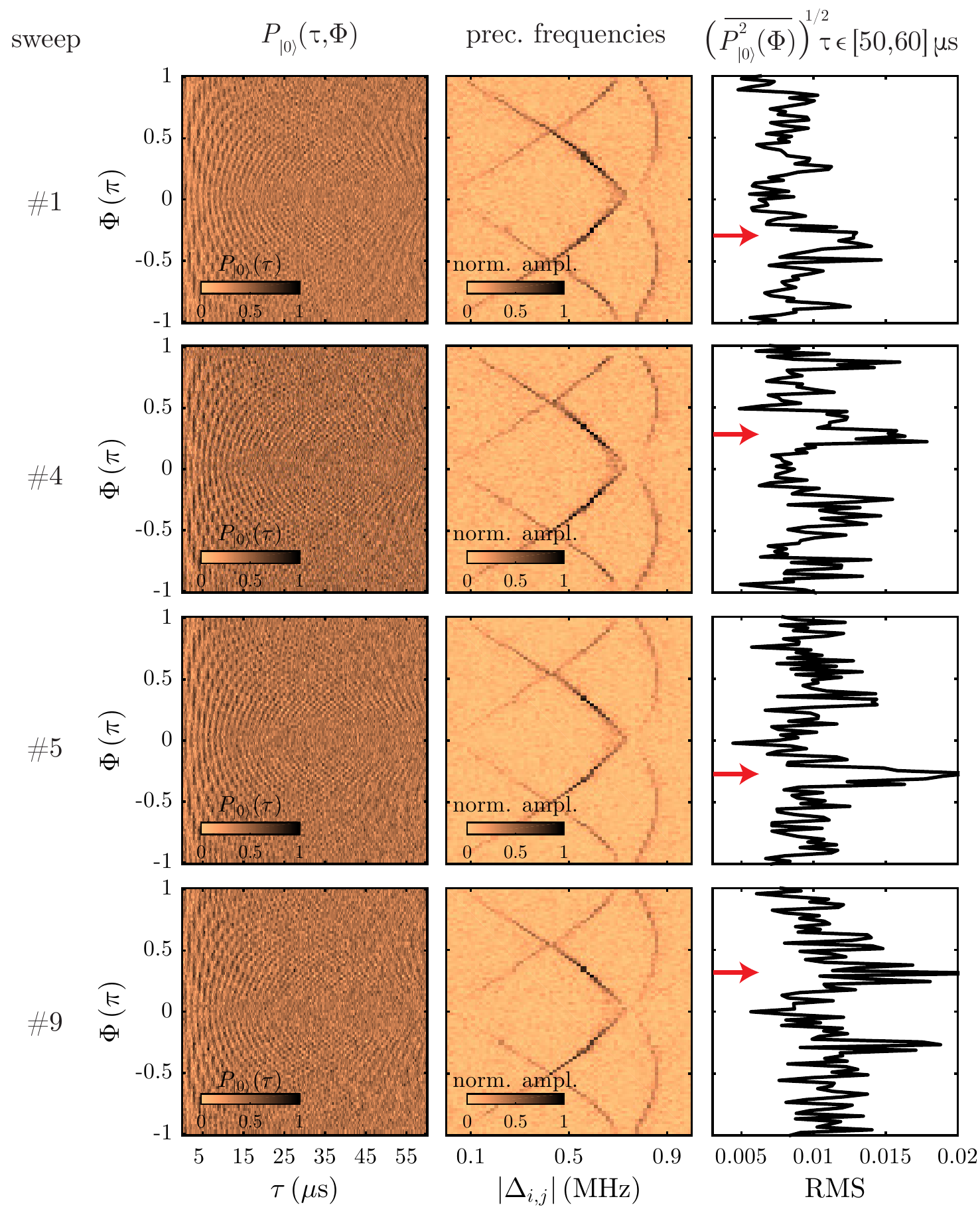}
	\caption{Time evolution of our experimental data. When investigating the time evolution of our closed-contour interaction scheme we essentially average over individually taken phase sweeps. Spin dynamics $P_{|0\rangle}(\tau,\Phi)$ (left), precession frequencies $\Delta_{i,j}$ (middle) and the root-mean square of $P_{|0\rangle}(\tau)$ for $\tau \in [50,60]$\,$\mu$s are shown here for four such sweeps. RMS maxima, indicating weak damping and marked by red arrows, change position with measurement time and thus hint at the presence of slowly evolving fluctuations.}
	\label{fig:App_QuantumBeat_NoiseSources_IndSweeps_RMS_Jan2017}
\end{figure}

To investigate the influence of zero-field splitting variations, represented by detunings $\delta D_0$ in our model (see Eq.~\eqref{eq:App_QuantumBeat_NoiseSources_Detunings}), on spin precession frequency spectra and Rabi decay time $T^\mathrm{dec}$, we repeated our simulations with the same noise environment, but non-zero detunings $\delta D_0$, and extracted precession frequencies $\Delta_{i,j}$ and Rabi decay times $T^\mathrm{dec}$ as described above. The results are shown in Fig.\,\ref{fig:App_QuantumBeat_NoiseSources_Overview_Comparions_InfluenceZFS} and indicate the following:
\begin{itemize}
	\item A fixed $\delta D_0/2\pi = -100$\,kHz (Fig.\,\ref{fig:App_QuantumBeat_NoiseSources_Overview_Comparions_InfluenceZFS}a) increases avoided crossings and induces a strong asymmetry in $T^\mathrm{dec}(\Phi)$. Specifically, the frequency component $\Delta_{+1,-1}$ at $\Phi \approx \pm 3\pi/4$ decays with $T^\mathrm{dec}_{+1,-1} \approx 120$\,$\mu$s, and therefore decoheres much more slowly than the $\Delta_{\pm1,0}$ component at $\Phi \approx \pm \pi/4$, which is characterized by  $T^\mathrm{dec}_{\pm 1,0} \approx 65$\,$\mu$s. 
	\item Decreasing $D_0$, i.e.\,setting $\delta D_0/2\pi = +100$\,kHz, also increases avoided crossings in the spin precession frequency spectrum. Inverting the polarity of the detuning, however, also inverts its effect on $T^\mathrm{dec}(\Phi)$. Now, the $\Delta_{\pm1,0}$ components decay with $T^\mathrm{dec}_{\pm 1,0} \approx 120$\,$\mu$s and the frequency components $\Delta_{+1,-1}$ disappear on timescales $T^\mathrm{dec}_{+1,-1} \approx 65$\,$\mu$s, 
\end{itemize}

By varying $\delta D_0$ with $\Phi$, i.e. with time during an experimental sweep of $\Phi$, (see inset to Fig.\,\ref{fig:App_QuantumBeat_NoiseSources_Overview_Comparions_InfluenceZFS}c) while solving the time-dependent Schr\"odinger equation, we can qualitatively reproduce the experimentally determined phase dependence of $T^\mathrm{dec}$ very well (compare Fig.\,\ref{fig:App_QuantumBeat_NoiseSources_Overview_Comparions_InfluenceZFS}c and Fig.\,\ref{fig:App_QuantumBeat_NoiseSources_Overview_Comparions_FFTTdecay}b). The employed variations in zero-field splitting $D_0$ of $\delta D_0/2\pi \approx \pm 100$\,kHz are in good agreement with the experimentally determined fluctuations (see Fig.\,\ref{fig:App_QuantumBeat_NoiseSources_Overview}c), and are most likely caused by environmental temperature fluctuations \cite{Acosta2010}. 
Note that we did not directly measure how $D_0$ actually evolved during our experiment, but rather infer the temporal variation of $D_0$ so as to find the best agreement between data and simulation. We attribute the remaining mismatch between experiment and theory to our lack of knowledge about the precise noise environment, especially slow fluctuations in driving field amplitudes and zero-field splitting, during the measurement.
The results from Fig.\,\ref{fig:App_QuantumBeat_NoiseSources_Overview_Comparions_InfluenceZFS}c should not be considered as fits, but rather a demonstration that including noise and drifts allows us to accurately reproduce our experimental observations through simulations.    

Our simulations indicate that slow fluctuations are responsible for the observed asymmetry in $T^\mathrm{dec}$ with respect to $\Phi$. We can support this statement by further analysis of our experimental data. Typically, $P_{|0\rangle}(\tau,\Phi)$ is obtained by averaging several complete phase sweeps over a total measurement duration of approximately 240~hours.
In Fig.\,\ref{fig:App_QuantumBeat_NoiseSources_IndSweeps_RMS_Jan2017} we plot $P_{|0\rangle}(\tau)$ (left), spin precession frequencies $\Delta_{i,j}$ (middle) and the root-mean square (RMS) of $P_{|0\rangle}(\tau)$ for $\tau \in [50,60]$\,$\mu$s (right) versus $\Phi$ for four of these phase sweeps (note that the data set used here is not the one from the main text, but was recorded under similar conditions). The root-mean square of  $P_{|0\rangle}(\tau)$ serves as a measure for the remaining contrast of spin precession and therefore corresponds to the decay time $T^\mathrm{dec}$ (extracting $T^{\textrm{dec}}$ by fitting was not possible with these data due to limited signal-to-noise ratio). which could not be extracted by fitting due to the limited signal-to-noise ratio of the experimental data. Whenever the RMS of $P_{|0\rangle}(\tau)$ is large, the decay time is long. One can see that the individual RMS spectra show different behaviors with respect to $\Phi$, i.e.\,the observable maxima differ in amplitude and slightly in position. This behavior is very similar to the influence of fluctuations in zero-field splitting $D_0$ (see discussion above) and thus confirms the presence of slow experimental fluctuations.

\section{Perturbative Calculations of Phase Fluctuations}

\label{sec:PertCalc}

\subsection{Overview}

In addition to numerically modeling the effects of noise, we also used perturbative techniques to derive an analytical expression for the global phase dependence of the coherence times, to better understand how to improve coherence times further. We consider a total Hamiltonian $\hat{\boldsymbol{H}} = \hat{\boldsymbol{H}}_0 + \hat{\boldsymbol{H}}^\prime(\tau)$ consisting of the unperturbed Hamiltonian of Eq.~\ref{eq:App_QuantumBeat_DerivationHamiltonian_RWAasintext} and a time dependent perturbation. Under symmetric resonant driving,

\begin{eqnarray}
\hat{\boldsymbol{H}}_0 = \frac{\hbar}{2} \begin{pmatrix} 
\delta_1 & \Omega & \Omega e^{i\Phi} \\
\Omega &  0 & \Omega \\ 
\Omega e^{-i\Phi} & \Omega & \delta_2
\end{pmatrix}
\label{equ:TotHam}
\end{eqnarray}

\noindent in the rotating frame, where $\delta_{1,2}$ could be due to, e.g., a small error in the frequency of the driving fields or a detuning due to the drift of the ZFS. As discussed in the previous section and as often noted in the literature \cite{Jamonneau2016}, the NV coherence time $T_2^*$ is often limited by magnetic field noise. We will therefore consider a perturbation Hamiltonian corresponding to magnetic field noise given by $\hat{\boldsymbol{H}}^\prime = g_{\mathrm{NV}} \mu _B B_z(\tau) \hat{S}_z$, where $g_{\mathrm{NV}} = 2.0028$ is the NV g-factor \cite{Loubser1977, Loubser1978}, $\mu _B$ is the Bohr magneton, $B_z(\tau)$ is a fluctuating magnetic field along the NV axis, and $\hat{S}_z$ is the spin-1 operator along the NV axis.

To find the effect of the perturbations, we calculate the corrections to the propagator $\hat{U}$ in the interaction picture~\cite{Aharon2016}:

\begin{eqnarray}
\hat{U}_I (\tau) &=& 1 + \hat{U}_I^{(1)} + \hat{U}_I^{(2)} \nonumber \\
&=& 1 + \frac{1}{i \hbar} \int_{0}^{\tau}\!\hat{H}^\prime_I (\tau ^\prime)\,\mathrm{d}\tau ^\prime + \frac{1}{(i \hbar)^2} \int_{0}^{\tau}\!\int_{0}^{\tau ^\prime} \! \hat{H}^\prime_I (\tau ^\prime)\hat{H}^\prime_I (\tau ^{\prime \prime})\,\mathrm{d}\tau ^{\prime \prime}\mathrm{d}\tau ^{\prime},
\label{equ:Propagator}
\end{eqnarray}

\noindent where the perturbation Hamiltonian in the interaction picture is given by 

\begin{eqnarray}
\hat{H}_I^\prime (\tau) = e^{i \hat{H}_0 \tau /\hbar} \hat{H}^\prime (\tau) e^{-i \hat{H}_0 \tau /\hbar}.
\label{equ:IntPert}
\end{eqnarray}

\noindent Using this expression for the perturbation Hamiltonian, we will calculate the transition probabilities and phase fluctuations caused by the perturbation.

\subsection{Transition Probabilities Between Eigenstates}

The first order corrections primarily affect the populations and are related to the transition rates between eigenstates.

If we prepare our system in one of the driven eigenstates $\lvert \Psi_i \rangle$, the first order transition probability to eigenstate $\lvert \Psi_j \rangle$ can be written succinctly as~\cite{Aharon2016}

\begin{eqnarray}
\lvert \langle \Psi_j \rvert \hat{U}_I^{(1)}(\tau) \lvert \Psi_i \rangle \rvert ^2 = \frac{\pi}{4 \sqrt{3} \hbar ^2} \, \mu_B ^2 \, g^2 \, \tau \, S_{\mathrm{noise}} \left( \Delta_{i,j}(\Phi) \right) \, (1 - \delta_{i,j}), \qquad i,j \in \lbrace -1,0,1 \rbrace ,
\label{equ:TransProb}
\end{eqnarray}

\noindent where $\Delta_{i,j}(\Phi) = \left( E_j(\Phi) - E_i(\Phi) \right)/h$ is the frequency difference between the initial and final states, $S_{\mathrm{noise}}(\Delta_{i,j}(\Phi))$ is the noise power spectral density at the frequency $\Delta_{i,j}(\Phi)$, and $\delta_{i,j}$ is the Kronecker delta.

Due to the dependence of $\Delta_{i,j}$ on the global phase $\Phi$, the spectral region to which the spin is sensitive can be tuned by changing the global phase. With the driving strength of $\Omega/2 \pi = 500\,\mathrm{kHz}$ we used in our experiments, this corresponds to a frequency range of $\Delta_{\mathrm{min}} = 0$ to $\Delta_{\mathrm{max}} = 800\,\mathrm{kHz}$. Stronger driving strengths will lead to larger dynamic ranges. This suggests we can use the dressed states to develop novel, $\Phi$-dependent relaxometry techniques.

\subsection{Effect of Noise on Coherence Times}

If we prepare the system in one of the dressed states $\lvert \Psi_i \rangle$, the phase $\varphi_i$ accrued by $\lvert \Psi_i \rangle$ to second order in the interaction picture under the influence of the perturbation $\hat{H}'$ is given by~\cite{Aharon2016}

\begin{align}
\tan \left( \varphi_i \right) =& \frac{\mathrm{Im} \lbrace \langle \Psi_i \rvert \hat{U}_I^{(1)}(\tau) \lvert \Psi_i \rangle + \langle \Psi_i \rvert \hat{U}_I^{(2)}(\tau) \lvert \Psi_i \rangle \rbrace}{\mathrm{Re} \lbrace 1 + \langle \Psi_i \rvert \hat{U}_I^{(1)}(\tau) \lvert \Psi_i \rangle + \langle \Psi_i \rvert \hat{U}_I^{(2)}(\tau) \lvert \Psi_i \rangle \rbrace} \nonumber \\ =& \frac{\mathrm{Im} \lbrace \langle \Psi_i \rvert \hat{U}_I^{(2)}(\tau) \lvert \Psi_i \rangle \rbrace}{\mathrm{Re} \lbrace 1 + \langle \Psi_i \rvert \hat{U}_I^{(2)}(\tau) \lvert \Psi_i \rangle \rbrace}
\label{TotalPhase}
\end{align}

\noindent where we have used the fact that for our system $\langle \Psi_i \rvert \hat{U}_I^{(1)}(\tau) \lvert \Psi_i \rangle = 0$, i.e. there is no first-order contribution to the phase in this case. If the perturbation is small, we can also simplify the denominator as $\mathrm{Re} \lbrace 1 + \langle \Psi_i \rvert \hat{U}_I^{(2)} \lvert \Psi_i \rangle \rbrace \approx 1$ and use the approximation $\tan \left( \varphi_i \right) \approx \varphi_i$. Putting these approximations all together, we arrive at a simple expression for the additional phase due to the magnetic field noise:

\begin{eqnarray}
\varphi_i \approx \mathrm{Im} \lbrace \langle \Psi_i \rvert \hat{U}_I^{(2)}(\tau) \lvert \Psi_i \rangle \rbrace.
\label{equ:SimplePhase}
\end{eqnarray}

\begin{figure}[h!]
	\centering
	\includegraphics[width=0.8\textwidth]{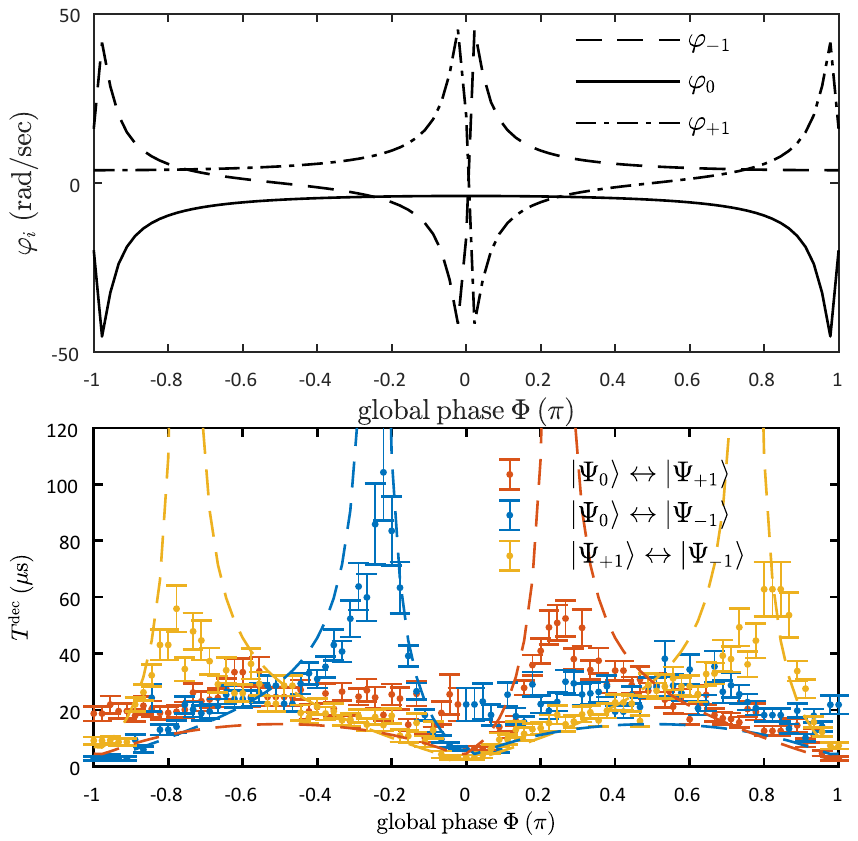}
	\caption{(a) The rate of phase pickup of the individual eigenstates under the influence of magnetic field noise. Note that the rates of phase acquisition for the $\lvert \Phi_{-1} \rangle$ and $\lvert \Psi_{+1} \rangle$ states, for example, are closest at $\Phi \approx \pm 0.75 \pi$. Since their rates of phase acquisition are similar, a two-level system spanned by these two states has a minimal dephasing rate as this value of $\Phi$, leading to longer coherence times at $\Phi \approx \pm 0.75 \pi$. (b) The values of $T^\mathrm{dec}_{i,j}$ extracted from the experimental data, as a function of $\Phi$, plotted alongside our analytical expressions.}
	\label{fig:CohTimes}
\end{figure}

There are several integral terms involved in calculating $\varphi_i$, but most of them average approximately to zero. Neglecting these terms and treating the magnetic field as an Ornstein-Uhlenbeck process $B_{\mathrm{O-U}}$, we find that the phase correction for $\lvert \Psi_{-1} \rangle$, for example, is given by

\begin{align}
\varphi_{-1} \approx \frac{g^2 \mu _B^2}{3 \hbar ^2} \, \tau \, \overline{B_{\mathrm{O-U}}^2 (\tau)} \Bigg( &{} \frac{E_{-1}(\Phi) - E_{+1}(\Phi)}{(1 / \tau_c)^2 + (E_{-1}(\Phi) - E_{+1}(\Phi))^2} +  \label{equ:m1Correction} \\
&{} \frac{E_{-1}(\Phi) - E_{0}(\Phi)}{(1 / \tau_c)^2 +(E_{-1}(\Phi) - E_{0}(\Phi))^2} \Bigg) \nonumber
\end{align}

\noindent where $ \overline{B_{\mathrm{O-U}}^2 (\tau)}= \frac{1}{\tau} \int_{0}^{\tau} \! B_{\mathrm{O-U}}^2 (\tau) \, \mathrm{d}\tau$ is the mean square value of $B_{\mathrm{O-U}}$ during the evolution time $\tau$ and $\tau_c$ is the noise correlation time \cite{Aharon2016}. Note that we have explicitly indicated the $\Phi$-dependence of the eigenenergies, to indicate that the amount of phase acquired depends on $\Phi$.
Expressions for $\varphi_{+1}$ and $\varphi_{0}$ can be found by permuting the indices.
Notice that $\varphi_{-1} = 0$ at $\Phi = -\pi/2$, while $\varphi_{+1} = 0$ at $\Phi = + \pi/2$ (see Fig. \ref{fig:CohTimes}), so that at certain values of $\Phi$, $\lvert \Psi_{+1} \rangle$ and $\lvert \Psi_{-1} \rangle$ are unaffected to second order by the presence of magnetic field noise along the NV axis.

For the initial state $\lvert 0 \rangle = \frac{1}{\sqrt{3}} \left( \lvert \Psi_{-1} \rangle + \lvert \Psi_0 \rangle + \lvert \Psi_{+1} \rangle \right)$, the phase accrued by one eigenstate $\lvert \Psi_i \rangle$ is given by

\begin{eqnarray}
\varphi_i \approx \mathrm{Im} \lbrace \langle \Psi_i \rvert \hat{U}_I^{(1)}(\tau) \lvert 0 \rangle + \langle \Psi_i \rvert \hat{U}_I^{(2)}(\tau) \lvert 0 \rangle \rbrace,
\label{equ:SimplePhase0}
\end{eqnarray}

\noindent which in principle involves the calculation of several more terms than in Eq.~\ref{equ:SimplePhase}. Those terms (both in first and second order) also approximately average to zero, so that Eq.~\ref{equ:m1Correction} (and the corresponding equations for $\varphi _0$ and $\varphi _{+1}$) still provides a decent approximation of the phase acquisition.

The coherence time is determined by the {\textit{relative}} phase acquisition between two eigenstates, such that the coherence time of the two-level system spanned by $\Psi_{-1}$ and $\Psi_{+1}$ (i.e., the $\Delta_{-1,+1}$ frequency component) is determined by the relative dephasing rate $\Gamma^{\mathrm{dec}}_{i,j} = \lvert \Delta \varphi_{i,j} \vert / \tau= \lvert \varphi_{i} - \varphi_{j} \rvert / \tau$, which is approximately given by

\begin{align}
\Gamma^{\mathrm{dec}}_{-1,+1} \approx \: &{} \frac{g^2 \, \mu _B^2}{3 \hbar ^2} \, \overline{B_\mathrm{O-U}^2(\tau)} \: \Bigg \vert \: 2 \frac{E_{-1}(\Phi) - E_{+1}(\Phi)}{(1/\tau _c)^2 + (E_{-1}(\Phi) - E_{+1}(\Phi))^2} + \label{equ:p1m1RelPhase} \\[2ex]
{}& \frac{E_{-1}(\Phi) - E_0 (\Phi)}{(1/\tau _c)^2+(E_{-1}(\Phi) - E_0(\Phi))^2} - \frac{E_{+1}(\Phi) - E_0(\Phi)}{(1/\tau _c)^2 + (E_{-1}(\Phi) - E_{0}(\Phi))^2} \: \Bigg \vert \nonumber.
\end{align}

\noindent Equivalent formulas for the other dephasing rates can be found by permuting the indices.
For Gaussian distributed phase fluctuations, the dephasing time $T_2 ^*$ is defined as the time it takes for $\overline{(\Delta \varphi)^2} = 2$ \cite{Jamonneau2016}. This yields an analytic expression for $T^{\mathrm{dec}} _{i,j}$:

\begin{align}
T^{\mathrm{dec}}_{-1,+1} = {}& \frac{\sqrt{2}}{\Gamma^{\mathrm{dec}}_{-1,+1}} \approx \frac{\sqrt{2} \, 3 \hbar^2}{g^2 \mu _B^2 \overline{B_\mathrm{O-U}^2 (\tau)}} \: \Bigg \vert \: 2 \frac{E_{-1}(\Phi) - E_{+1}(\Phi)}{(1/\tau _c)^2 + (E_{-1}(\Phi) - E_{+1}(\Phi))^2} + \label{equ:p1m1CohTime} \\[2ex]
{}& \frac{E_{-1}(\Phi) - E_0 (\Phi)}{(1/\tau _c)^2+(E_{-1}(\Phi) - E_0(\Phi))^2} - \frac{E_{+1}(\Phi) - E_0(\Phi)}{(1/\tau _c)^2 + (E_{-1}(\Phi) - E_{0}(\Phi))^2} \: \Bigg \vert ^{-1} \nonumber,
\end{align}

\noindent where the expressions for the other Rabi decay times can be found by permuting the indices. To compare our analytical equations to the simulations in the first part of our supplementary material, we use the values $\Omega / 2 \pi = 500 \, \mathrm{kHz}$, $\delta_1 / 2 \pi = \delta_2 /2 \pi = 10 \, \mathrm{kHz}$, $\tau_c = 10 \, \mathrm{\mu s}$, $\sqrt{\overline{B_{\mathrm{O-U}}^2 (\tau)}} = 3.8 \, \mathrm{\mu T}$ (corresponding to $T_2 ^* = 2.1 \, \mathrm{\mu s}$), to estimate and plot the expressions for all three $T^\mathrm{dec} _{i,j}$ alongside the data in Fig.~\ref{fig:CohTimes}b, which shows qualitative agreement between the data and the expressions.

The approximate expression in Eq.~\ref{equ:p1m1CohTime} accurately predicts the asymmetric shape around the maxima, though the positions of the maxima are slightly off.
Also note that the analytic expressions in Eq.~\ref{equ:p1m1CohTime} diverge for certain values of $\Phi$.
In addition to the relatively quickly changing magnetic field $B_{\mathrm{O-U}}$, however, we must also account for very slowly changing magnetic fields, which are constant during a single measurement or evolution time but can change between subsequent ones; these constant fields have the effect of changing the values of $\delta_1 , \delta_2$ in the unperturbed Hamiltonian Eq.~\ref{equ:TotHam}, with the condition $\delta^B \equiv \delta^{B}_1 = - \delta^{B}_2$.
We include the effect of slowly varying fields by integrating over the possible dephasing rates as a function of the induced detuning, weighted by the probability of that detuning:

\begin{align}
\Gamma^{\mathrm{dec}}_{-1,+1} (\Phi) = \int^{\infty}_{- \infty} \! P(\delta^B) \Gamma^{\mathrm{dec}}_{-1,+1} (\Phi,\delta^B) \, \mathrm{d} \delta^B,
\end{align}

\noindent where $\Gamma ^{\mathrm{dec}} _{-1,+1} (\Phi , \delta^B)$ is the dephasing rate in Eq.~\ref{equ:p1m1RelPhase} (where we now treat the eigenergies $E_{k}$ as functions of $\delta^B$) and $P ( \delta^B)$ is the probability distribution of the detunings caused by the slowly varying fields. For $P(\delta^B)$, we use a normal distribution with mean $\mu = 0$ and variance $\sigma^2 = (107 \, \mathrm{kHz})^2$, corresponding to $T_2^* = 2.1 \, \mathrm{\mu s}$.
Performing this integral removes the divergence and leads to a maximum decoherence time on the order of $1 \, \mathrm{ms}$.
Other discrepancies between Eq.~\ref{equ:p1m1CohTime} and the data are likely due to the contributions of other noise sources (such as fluctuations in the zero field splitting or noise in the strength of the driving fields), which will also have the effect of reducing the maximum achievable coherence time. Even so, the simple expression in Eq.~\ref{equ:p1m1CohTime} captures the functional form and scale of the coherence times, simply by entering realistic values for the parameters.

Our perturbative calculations show that the improved coherence times are due to the fact that for certain values of $\Phi$ two eigenstates experience the same amount of phase acquisition, such that their relative phase is unchanged by the presence of the magnetic noise.
Furthermore, Eq.~\ref{equ:p1m1CohTime} shows the intuitively obvious result that stronger driving leads to improved coherence times, since stronger driving leads to larger energy splittings.
Finally, the similarity between the data and the calculations confirms that the bulk of the noise in our system is due to fluctuating magnetic fields projected along the NV axis.
Thus, removing excess spins from our sample would prolong coherence times and change the $\Phi$-dependence of the coherence times, as other sources of noise come to dominate.

\section{Creation of phase-locked driving fields}
\label{sec:App_QuantumBeat_CreationDrivingFields}

\begin{figure}[!h]
	\centering
	\includegraphics[]{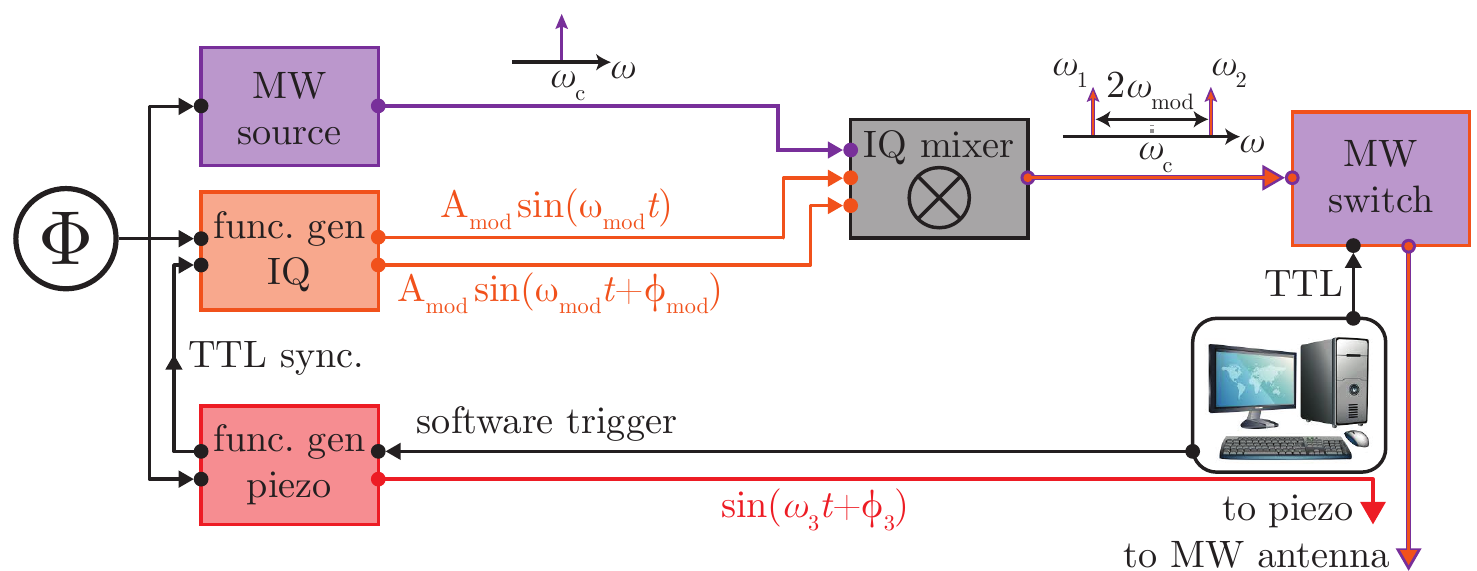}
	\caption{Creation of driving fields for closed-contour interaction. The two MW tones with frequencies $\omega_{1,2} = \omega_\mathrm{c} \pm \omega_\mathrm{mod}$ are created through frequency modulating a carrier signal of frequency $\omega_\mathrm{c}$ at frequency $\omega_\mathrm{mod} = \omega_3/2$. Phase-locking of the three driving fields is achieved via pulsed output synchronization and locking of MW source, IQ and piezo function generators to the same 10\,MHz reference signal.}
	\label{fig:App_StrongDriving_ExplanationCrossings_Circuit}
\end{figure}

We create the two MW tones used to drive the $|0\rangle \leftrightarrow |\pm1\rangle$ transitions by frequency modulating a carrier signal $S_\mathrm{c}$ at frequency $\omega_\mathrm{c}$ and amplitude $A_\mathrm{c}$ with two time-dependent modulation signals
\begin{equation}
S_\mathrm{mod,I}(t) = A_\mathrm{mod} \sin\left( \omega_\mathrm{mod} t + \phi_\mathrm{mod} \right)
\label{eq:App_QuantumBeat_CreationDrivingFields_Imod}
\end{equation}
and
\begin{equation}
S_\mathrm{mod,Q}(t) = A_\mathrm{mod} \sin\left( \omega_\mathrm{mod} t  \right)
\label{eq:App_QuantumBeat_CreationDrivingFields_Qmod}
\end{equation}
of equal, but constant amplitudes $A_\mathrm{mod}$. These signals are mixed to the carrier signal as $I$ and $Q$ modulation inputs, and we obtain the output signal
\begin{equation}
\begin{split}
S_\mathrm{out}\left(t\right) & = S_\mathrm{mod,I}(t) \mathrm{Re} \lbrace S_\mathrm{c} \rbrace +  S_\mathrm{mod,Q}(t) \mathrm{Im} \lbrace S_\mathrm{c} \rbrace\\
&= A_\mathrm{c} A_\mathrm{mod} \left[ \sin\left( \frac{\phi_\mathrm{mod} - \pi/2}{2}\right) \cdot \cos\left( \left(\omega_\mathrm{c} + \omega_\mathrm{mod}\right)t + \frac{\pi/2 + \phi_\mathrm{mod}}{2}\right)\right]\\ 
&\,\,\,\,\,\,+ A_\mathrm{c} A_\mathrm{mod} \left[ \sin\left( \frac{\phi_\mathrm{mod} + \pi/2}{2}\right) \cdot \cos\left( \left(\omega_\mathrm{c}-\omega_\mathrm{mod}\right)t + \frac{\pi/2-\phi_\mathrm{mod}}{2}\right)\right],
\end{split}
\label{eq:App_QuantumBeat_CreationDrivingFields_IQOutput}
\end{equation}
which consists of two MW tones separated by $2\omega_\mathrm{mod}$ and symmetrically located around $\omega_\mathrm{c}$. 
The relative phase $\phi_\mathrm{mod}$ of the two modulation signals allows for modifying the relative amplitudes of the two MW tones. This is usually necessary to establish the condition $\Omega_1 = \Omega_2$, as our MW antenna does not deliver a fully linearly polarized MW field to the NV centre.

As demonstrated earlier, we can only define a time-independent global phase $\Phi$ if the closed-contour condition $\omega_1 + \omega_3 = \omega_2$ is fulfilled. To ensure that this is always the case we choose $\omega_\mathrm{c}/2\pi = D_0$ and $\omega_\mathrm{mod} = \omega_3/2$ with $\omega_3$ being the eigenfrequency of our mechanical resonator. 
The global phase $\Phi = \phi_1 + \phi_3 - \phi_2$ becomes
\begin{equation}
\Phi = \phi_3 - \left(\phi_\mathrm{mod} + \pi\right)
\label{App_QuantumBeat_CreationDrivingFields_GlobalPhase}
\end{equation}
under such conditions and for $0 \leq \phi_\mathrm{mod} < \pi/2$. The global phase $\Phi$ can therefore be controlled by changing the individual phase $\phi_3$ of the sinusoidal signal that drives the mechanical actuation of our diamond resonator.

To create the phase-locked driving fields experimentally, we connect the MW generator (Stanford Research Systems, SRS384), the function generator driving the piezo for mechanical actuation (Keysight, 33522A) and the function generator that supplies the IQ modulation signals (Keysight, 33622B) to the same 10\,MHz reference signal. To set the global phase to a reproducible value, the output of the piezo function generator is triggered via a software command. Upon receiving the software trigger, it emits another trigger pulse which starts the output of our IQ modulation function generator (Fig.\,\ref{fig:App_StrongDriving_ExplanationCrossings_Circuit}). During our experiment mechanical actuation of our resonator is always active and MW pulses are created by employing a MW switch (MiniCircuits, ZASWA-2-50DR+) with a rise-time of 2\,ns, controlled via digital pulses from our fast pulse generator card.

\end{document}